\documentclass[
superscriptaddress,
 amsmath,amssymb,
 aps,
 pra,
 10pt,
 twocolumn,
 longbibliography,
 floatfix,
]{revtex4-2}

\usepackage{graphicx}
\usepackage{dcolumn}
\usepackage{bm}
\usepackage{bbm}
\usepackage{braket}
\usepackage{physics}
\usepackage{amsmath,amssymb,amsthm}
\usepackage{xcolor}
\usepackage[colorlinks=true,citecolor=blue,linkcolor=blue,urlcolor=blue]{hyperref}
\usepackage{mathrsfs}
\usepackage{scalefnt}

\newcommand{\Emph}[1]{\emph{\textbf{#1.}}~---}

\newcommand{\id}{\mathbbm{1}}
\renewcommand{\vec}[1]{{\boldsymbol{#1}}}
\newcommand{\expect}[1]{\langle #1\rangle}
\newcommand{\ud}{\mathrm{d}}

\newcommand{\mc}[1]{\mathcal{#1}}

\renewcommand{\L}{\mc{L}}
\newcommand{\D}{\mc{D}}
\newcommand{\mri}{\mathrm{i}\mkern1mu}

\newcommand{\ha}{\hat{a}}
\newcommand{\hn}{\hat{n}}
\newcommand{\hw}{\hat{w}}
\newcommand{\hA}{\hat{A}}
\newcommand{\hB}{\hat{B}}
\newcommand{\hH}{\hat{H}}
\newcommand{\hL}{\hat{L}}
\newcommand{\hM}{\hat{M}}
\newcommand{\hR}{\hat{R}}
\newcommand{\hS}{\hat{S}}
\newcommand{\hU}{\hat{U}}
\newcommand{\hV}{\hat{V}}
\newcommand{\hsigma}{\hat{\sigma}}
\newcommand{\dm}{\hat{\rho}}
\newcommand{\dmf}{\hat{\varrho}}
\newcommand{\up}{{\uparrow}}
\newcommand{\down}{{\downarrow}}
\newcommand{\Right}{{\rightarrow}}
\newcommand{\Left}{{\leftarrow}}

\newcommand{\dqc}{Duke Quantum Center, Duke University, Durham, NC 27701, USA}
\newcommand{\dukephysics}{Department of Physics, Duke University, Durham, NC 27708, USA}

\newcommand{\Title} {Non-Gaussian Phase Transition and Cascade of Instabilities in the Dissipative Quantum Rabi Model}
\newcommand{\Authors}
{
\author{Mingyu Kang}
\email{These authors contributed equally.}
\affiliation{\dqc}
\affiliation{\dukephysics}

\author{Yikang Zhang}
\email{These authors contributed equally.}
\affiliation{\dukephysics}
\affiliation{Department of Physics and Astronomy, Dartmouth College, Hanover, NH 03755, USA}

\author{Kenneth R. Brown}
\affiliation{\dqc}
\affiliation{\dukephysics}
\affiliation{Department of Electrical and Computer Engineering, Duke University, Durham, NC 27708, USA}
\affiliation{Department of Chemistry, Duke University, Durham, NC 27708, USA}

\author{Thomas Barthel}
\affiliation{\dqc}
\affiliation{\dukephysics}
\affiliation{National Quantum Laboratory, University of Maryland, College Park, MD 20742, USA}
\affiliation{Department of Physics, University of Maryland, College Park, MD 20742, USA}
}
\newcommand{\Date} {January 8, 2026}

\begin{document}

\title{\texorpdfstring{Non-Gaussian Phase Transition and Cascade of Instabilities\\ in the Dissipative Quantum Rabi Model}{\Title}}
\Authors
\date{\Date}

\begin{abstract}
The open quantum Rabi model describes a two-level system coupled to a harmonic oscillator. A Gaussian phase transition for the nonequilibrium steady states has been predicted when the bosonic mode is soft and subject to damping. We show that oscillator dephasing is a relevant perturbation, which leads to a non-Gaussian phase transition and an intriguing cascade of instabilities for $k$th-order bosonic operators, as well as a jump in the steady-state qubit polarization. For the soft-mode limit, the equations of motion form a closed hierarchy and spectral properties can be efficiently studied. To this purpose, we establish a fruitful connection to non-Hermitian Hamiltonians. The results for the phase diagram, stability boundaries, and relevant observables are based on mean-field analysis, exact diagonalization, perturbation theory, and Keldysh field theory.
\end{abstract}

\maketitle

Coupled spin-boson systems are ubiquitous in physics, describing atoms in optical cavities~\cite{jaynes1963comparison,zueco2009qubit,braak2011integrability,strack2011dicke,bhaseen2012dynamics}, superconducting qubits coupled to electromagnetic modes~\cite{Leggett87,tian2002decoherence,thorwart2004dynamics,gulacsi2023signatures}, electronic excitations in molecular systems or quantum dots interacting with electric dipole moments of the solvent and phonons~\cite{Garg85,Gilmore05,macdonell2021analog,Kang24}, and energy transfer in biological light-harvesting complexes subject to vibrations~\cite{Huelga13,Jang18,ko2022dynamics}. When a system parameter is varied in a closed system, one can encounter a nonanalytic change in the ground state, i.e., a quantum phase transition (QPT)~\cite{Sachdev,sondhi1997continuous}. Such QPTs also occur in spin-boson systems like the Rabi model~\cite{ashhab2013superradiance,hwang2015quantum,liu2017universal,ye2024superradiant} and the Dicke model~\cite{hepp1973superradiant,wang1973phase,emary2003chaos}, where the thermodynamic limit required for a QPT involves a divergence of the oscillator occupation number. Similarly, nonequilibrium steady states of open quantum systems that interact with environmental degrees of freedom can undergo nonanalytic changes~\cite{hwang2018dissipative,de2023signatures,li2024spin,wu2024experimental,nagy2011critical,torre2013keldysh,kirton2017suppressing,gegg2018superradiant}. This phenomenon is called a dissipative phase transition (DPT)~\cite{kessler2012dissipative}. Signatures of QPT and DPT are experimentally observed in various systems, including optical cavities \cite{baumann2010dicke,brennecke2013real,baden2014realization}, trapped ions~\cite{cai2021observation}, and circuit quantum electrodynamics~\cite{zheng2023observation}.

The interactions between a quantum oscillator and its environment feature two prominent decoherence channels: damping and dephasing~\cite{memarzadeh2016minimum,michael2016new}.
Density operators $\dmf$ of Markovian open systems evolve according to Lindblad master equations~\cite{Lindblad1976-48,Gorini1976-17},
\begin{equation}\label{eq:introlind}
    \partial_t\dmf = \L(\dmf) = -\mri [\hH,\dmf]+\sum_\alpha\kappa_\alpha\D[\hL_\alpha](\dmf),
\end{equation}
with the Liouville superoperator $\L$, Hamiltonian $\hH$, and dissipators $\D[\hL_\alpha](\dmf):=\hL_\alpha\dmf\hL_\alpha^\dag-\frac{1}{2}\{\hL_\alpha^\dag\hL_\alpha,\dmf\}$. Bosonic damping and dephasing are described by the dissipators $\D[\ha]$ and $\D[\hn]$, respectively, where $\ha$ is the bosonic annihilation operator and $\hn:=\ha^\dag \ha$ the number operator.

Dissipative bosonic systems have been studied extensively. Without non-linear processes like dephasing, the system is quasi-free such that Gaussianity is preserved throughout the evolution~\cite{weedbrook2012gaussian,linowski2022dissipative}, and one can employ the third quantization technique~\cite{prosen2010quantization,Barthel2022} to diagonalize the Liouvillian $\L$. Nonlinear effects make the problem more challenging, and hence spin-boson systems with bosonic dephasing have been much less explored, although dephasing is, for example, known to be a dominant decoherence mechanism in molecular systems ~\cite{nitzan2006chemical,stock1990theory,banin1994impulsive,olaya2025simulating}. Dephasing also occurs naturally in superconducting microwave cavities and mechanical resonators whose resonance frequency $\omega_0$ jitters due to fluctuating electric fields from charge defects or vibrations in the environment \cite{Reagor2016-94,Leviant2022-6,Pietikainen2024-5}. This dephasing can be engineered and controlled. Thus, models with bosonic dephasing are promising candidates for studying \textit{non-Gaussian phase transitions}, i.e., nonanalytic changes in non-Gaussian steady states. 

\begin{figure}[t]
\includegraphics[width=\linewidth]{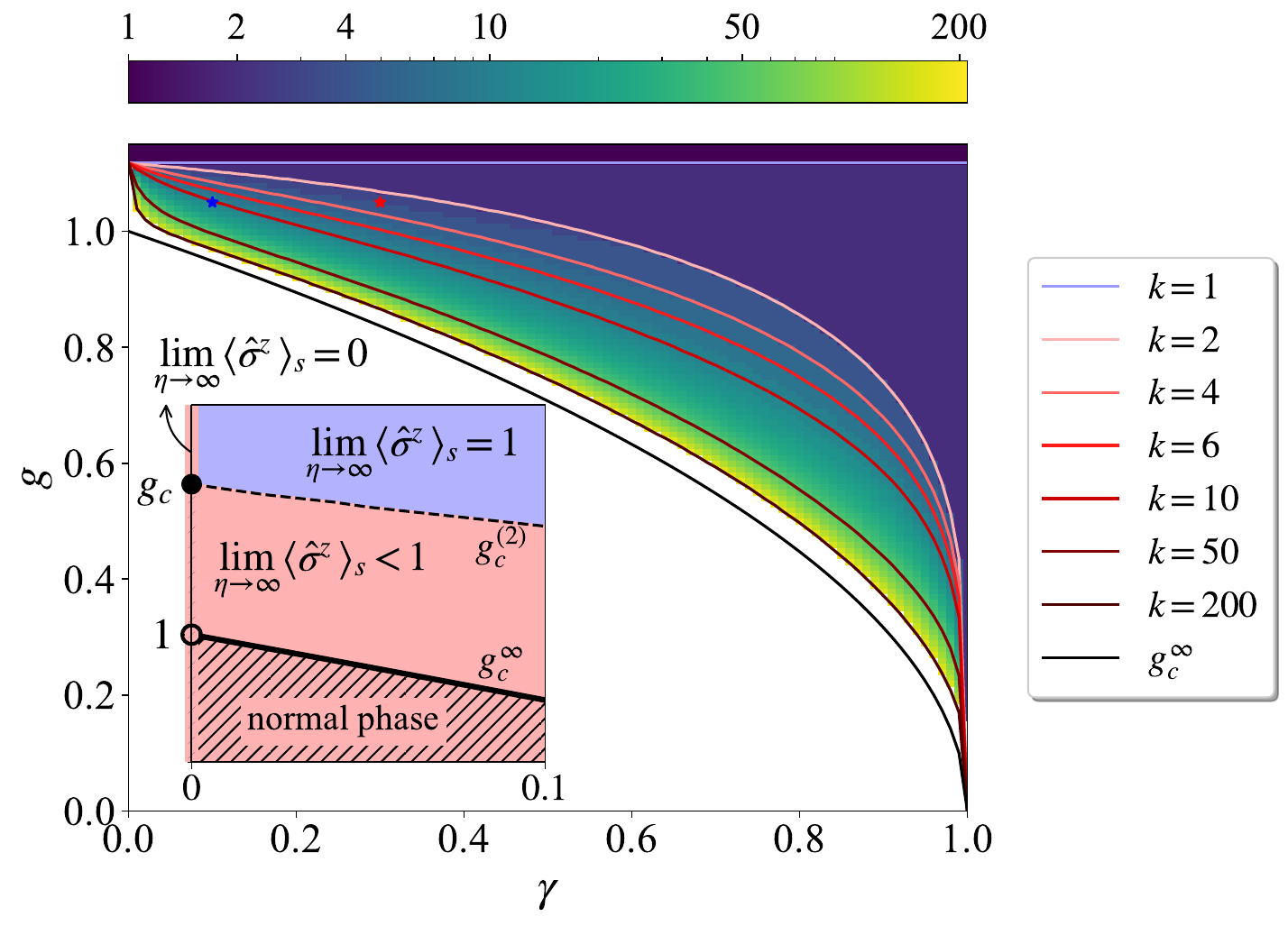}
	\caption{\label{fig:boundary} Stability boundaries of the $k$th-order bosonic operators in the thermodynamic limit $\eta \to \infty$. The color map indicates the smallest $k$ for which, at the given point in the phase diagram, the non-Hermitian Hamiltonian $\hH^{(k)}$ from Eq.~\eqref{eq:eff_Hk} is unstable (not a Hurwitz matrix). The black line indicates $g^\infty_c$ from Eq.~\eqref{eq:gcinfty}, below (above) which the whole system is stable (unstable). The star-shaped markers indicate the parameters used in Fig.~\ref{fig:hierarchy}. Here we set $\kappa/\omega_0 = 1$. \textbf{Inset:} Infinitesimal dephasing $\gamma>0$ causes nonanalytic jumps in the qubit polarization and the normal-phase boundary, leading to the non-Gaussian DPT at $g_c^\infty$.}
\end{figure}

In this work, we study the open quantum Rabi model, where a two-level system (qubit) interacts with a bosonic mode (oscillator), and the latter experiences damping and dephasing due to the environment. In the soft-mode limit where the qubit transition frequency $\Omega$ is much larger than the oscillator frequency $\omega_0$, we find that arbitrarily small oscillator dephasing drives the system away from the Gaussian DPT of the purely damped Rabi model described in Ref.~\cite{hwang2018dissipative} to a non-Gaussian DPT. We identify two distinctive features of the DPT (see Fig.~\ref{fig:boundary}): First, coming from the normal phase, higher-order operators diverge before lower-order operators, leading to a cascade of instabilities. Second, infinitesimal dephasing induces a nonanalytic jump in both the normal-phase boundary and the steady-state qubit polarization. These phenomena are explained by establishing a novel connection between the closed hierarchy of dynamic equations for the bosonic operators~\cite{unkovi2014, Barthel2022, zhang2022criticality} to non-Hermitian spin Hamiltonians, admitting for an efficient analysis of spectral properties.

\Emph{Model}
Consider the quantum Rabi model
\begin{equation}
    \hH_\text{Rabi} = \frac{\Omega}{2} \hsigma^z + \lambda \hsigma^x (\ha + \ha^\dag) + \omega_0 \hn,
\end{equation}
with the Pauli operators $\hsigma^r$, qubit transition frequency $\Omega$, oscillator frequency $\omega_0$, and coupling strength $\lambda$. Following Refs.~\cite{hwang2015quantum, hwang2018dissipative}, we define the frequency ratio and dimensionless coupling strength
\begin{equation}
    \eta := \Omega/\omega_0\quad\text{and}\quad
    g := 2\lambda/\sqrt{\omega_0 \Omega}.
\end{equation}
The oscillator is also subject to damping and dephasing such that the Lindblad master equation~\eqref{eq:introlind} for the qubit-oscillator density matrix $\dmf$ reads
\begin{equation}\label{eq:mastereqn}
    \partial_t\dmf = -\mri [\hH_\text{Rabi},\dmf]+\kappa(1-\gamma)\D[\ha](\dmf)+\kappa\gamma\D[\hn](\dmf).
\end{equation}
The dissipation is parametrized by the overall rate $\kappa$ and the dimensionless \emph{non-Gaussianity} parameter $\gamma$, which tunes from pure damping at $\gamma=0$ to pure dephasing at $\gamma=1$.
The model has a weak $\mathbb{Z}_2$ symmetry in the sense that $\L(\hV\hR\hV^\dag)=\hV\L(\hR)\hV^\dag$ for any operator $\hR$ and the symmetry transformation $\hV=\hsigma^z(-1)^{\hn}$, which causes a sign change for $\hsigma^x$, $\hsigma^y$, and $\ha$.

\Emph{Mean-field theory}
Decoupling operator products in the form $\expect{\hsigma^r\ha}\approx\expect{\hsigma^r}\expect{\ha}$, the mean-field analysis as detailed in Appx.~\ref{appx:MFT} suggests a DPT at coupling strength 
\begin{equation}\label{eq:MFT-gc}
	g_c:= \sqrt{1+\kappa^2/(4\omega_0^2)}
\end{equation}
with spontaneous $\mathbb{Z}_2$ symmetry breaking. The expectation value $\expect{\ha}$ is zero in the normal phase $g<g_c$ but, in the superradiant phase $g>g_c$, it diverges for the soft-mode limit $\eta\to\infty$, and we can define the order parameter $\expect{\ha}/\sqrt{\eta}$. Above the transition point, it is proportional to $\pm\sqrt{1-(g/g_c)^4}$. With the mean boson number becoming macroscopic while quantum fluctuations are suppressed, $\eta\to\infty$ is the thermodynamic limit~\cite{Carmichael2015,hwang2015quantum,Bartolo2016-94,Casteels2017-95,hwang2018dissipative,Zhang2021-103}.

Interestingly, in the Heisenberg picture,
\begin{equation}\label{eq:dampdephsame}
    \D^\dag[\ha](\ha) = \D^\dag[\hn](\ha) = -\ha/2
\end{equation}
where $\D^\dag[\hL](\hA) = \hL^\dag\hA\hL - \frac{1}{2}\{\hL^\dag \hL,\hA\}$ are the adjoint dissipators. In other words, the mean-field analysis cannot distinguish damping and dephasing. The mean-field equations of motion (EOM) are independent of $\gamma$, and the results coincide with the dephasing-free case analyzed in Ref.~\cite{hwang2018dissipative}. We will see in the following that, in reality, dephasing leads to qualitatively different physics.

\Emph{Soft-mode limit}
For large $\eta$, one can use the unitary Schrieffer-Wolff transformation \cite{Schrieffer1966-149} $\hU := \exp[-(g\sqrt{\eta}/2)(\ha+\ha^\dag)(\hsigma^+ - \hsigma^-)]$ to decouple the qubit from the bosonic mode to order $\eta^{-1/2}$ as discussed in Ref.~\cite{hwang2015quantum}. In the $\eta\to\infty$ limit, the transformed Hamiltonian reads (up to a constant)
\begin{equation}\label{eq:HafterSW}
	\hU^\dag \hH_\text{Rabi} \hU = \frac{\Omega}{2}\hsigma^z + \frac{\omega_0g^2}{4}\hsigma^z(\ha+\ha^\dag)^2+\omega_0\hn,
\end{equation}
where the terms involving the qubit are diagonal in the $\hsigma^z$ basis. The dissipators do not change under the transformation $\hU$ to order $\eta^{-1}$~\cite{hwang2018dissipative} and only act on the transformed bosonic mode. We then have a strong $\mathbb{Z}_2$ symmetry \cite{Buca2012,Albert2014} in the sense that the transformed Hamiltonian \eqref{eq:HafterSW} and Lindblad operators commute with $\hV=\hsigma^z(-1)^{\hn}$. This results in a two-dimensional steady-state manifold given by convex combinations of
\begin{align}\label{eq:degenerate}
	\dmf_{s,\down}=\ket{\down}\bra{\down}\otimes\dm_\down \qq{and}
	\dmf_{s,\up}=\ket{\up}\bra{\up}\otimes\dm_\up.
\end{align}
When the qubit is in the $\hsigma^z$-eigenstate $\ket{\down}$, the reduced Hamiltonian for the oscillator becomes
\begin{equation}\label{eq:Hdown}
	\hH_\down = -\frac{\omega_0g^2}{4}(\ha+\ha^\dag)^2+\omega_0\hn.
\end{equation}
Thus, the oscillator state $\dm_\down$ follows the Lindblad master equation $\partial_t\dm_\down = \L_\down(\dm_\down)$ with
\begin{equation}\label{eq:mastereqn2}
	\L_\down(\dm_\down) = -\mri [\hH_\down,\dm_\down]+\kappa(1-\gamma)\D[\ha](\dm_\down)+\kappa\gamma\D[\hn](\dm_\down).
\end{equation}
For the case with the qubit in the $\hsigma^z$-eigenstate $\ket{\up}$, one needs to substitute $g^2\to-g^2$ in Eq.~\eqref{eq:Hdown}. 
In the following, we analyze the stability for $\L_\down$. In fact, it determines the overall stability as shown by the analogous analysis of $\L_\up$ in Appx.~\ref{appx:odd-k_up}.

In the case of pure damping ($\gamma=0$), the steady state is a Gaussian state, which is fully characterized by its covariance matrix (the static 2-point Green's function) \cite{prosen2010quantization,Barthel2022}. All steady-state observables can then be obtained using Wick's theorem \cite{Wick1950-80,Negele1988}. However, in the presence of the dephasing channel $\D[\hn]$, the steady state is non-Gaussian.
Fortunately, as discussed in Refs. \cite{unkovi2014,Barthel2022,zhang2022criticality}, due to the Hermiticity of $\hn$, the dynamics is still solvable in the sense that the EOM for the bosonic $k$th-order operators $(\ha^\dag)^m \ha^{k-m}$ ($m=0,\dotsc,k$) form a closed hierarchy.
Specifically, the time derivatives of $k$th-order bosonic operators depend only on the bosonic operators of order $k$, $k-2$, $k-4$, etc.

\Emph{Instability cascade and non-Hermitian Hamiltonians}
Analyzing the Liouvillian \eqref{eq:mastereqn2}, we will find a cascade of stability boundaries $g_c^{(k)}$ as shown in Fig.~\ref{fig:boundary}. For coupling strengths $g>g_c^{(k)}$ the expectation values of $k$th-order operators are unstable in the sense that some of them diverge in the thermodynamic limit $\eta\to\infty$ but they are all stable for $g<g_c^{(k)}$. These boundaries are ordered such that
\begin{equation}
	g_c^{(k+2)}\leq g_c^{(k)}\leq g_c^{(1)}\equiv g_c
	\qq{for all} k,
\end{equation}
i.e., coming from the normal phase, higher-order observables diverge first. There are two separate cascades for even and odd $k$.

For small $k$, the EOM can be assessed directly. Specifically, for $k=1$, the vector $\vec{v}_1 := (\expect{\ha}, \expect{\ha^\dag})^T$ follows the EOM $\partial_t\vec{v}_1 = H_1\vec{v}_1$, where
\begin{equation}\label{eq:M1}
	H_1 = \mri\omega_0 \begin{pmatrix}
	\frac{\mri\kappa}{2\omega_0} - 1+\frac{g^2}{2} & \frac{g^2}{2} \\
	-\frac{g^2}{2} & \frac{\mri\kappa}{2\omega_0}+ 1-\frac{g^2}{2}
	\end{pmatrix},
\end{equation}
which has eigenvalues $\ell^{(1)}_\pm = -\kappa/2 \pm \mri\omega_0\sqrt{1-g^2}$. Thus, $\vec{v}_1$ is stable if and only if $g<g^{(1)}_c:= \sqrt{1 + \kappa^2/(4\omega_0^2)}$ such that the real parts of both eigenvalues $\ell^{(1)}_\pm$ are negative. Note that $g^{(1)}_c$ coincides with the mean-field value $g_c$ and does not depend on the non-Gaussianity parameter $\gamma$. This is because, according to Eq.~\eqref{eq:dampdephsame}, damping and dephasing have identical effects on the first-order bosonic operators.

The effects of non-Gaussianity are revealed by bosonic operators of order $k\geq 2$. The vector $\vec{v}_2 := (\expect{\ha\ha}, \expect{\ha^\dag\ha}, \expect{\ha^\dag\ha^\dag})^T$ follows the EOM $\partial_t\vec{v}_2 = H_2\vec{v}_2+Y_2$, where
\begin{equation*}\label{eq:M2}
	H_2 = \mri\omega_0 \begin{pmatrix}
	\frac{\mri(1+\gamma)\kappa}{\omega_0} - 2+g^2 & g^2 & 0 \\
	-\frac{g^2}{2} & \frac{\mri(1-\gamma)\kappa}{\omega_0} & \frac{g^2}{2} \\
	0 & -g^2 & \frac{\mri(1+\gamma)\kappa}{\omega_0} + 2-g^2
	\end{pmatrix}
\end{equation*}
and $Y_2 = \mri\omega_0(\frac{g^2}{2}, 0, -\frac{g^2}{2})^T$. Thus, $\vec{v}_2$ is stable if and only if all $H_2$ eigenvalues have negative real parts. Appendix~\ref{appx:k2} shows that this condition is equivalent to $g < g^{(2)}_c$ with
\begin{align}\label{eq:gc2}\textstyle
	g^{(2)}_c := \sqrt{\frac{\sqrt{1-\gamma}}{\gamma}
	 \Big(\sqrt{1+\gamma+\frac{\gamma\kappa^2}{2\omega_0^2}(1+\gamma)^2} - \sqrt{1-\gamma}\Big)
	 }.
\end{align}
Note that $\lim_{\gamma \to 0} g^{(2)}_c = g_c$.

The stability conditions for $k>2$ can be obtained similarly, and the computational effort for an exact determination is $\order{k^3}$: As shown in Appx.~\ref{appx:eff_H}, the generator for the EOM of the bosonic operators has a block-triangular structure, and the diagonal block $\hH^{(k)}$ for the $k$th-order operators can be written as an effective non-Hermitian Hamiltonian for a simple spin-$k/2$ model. Specifically, it takes the form
\begin{equation}\label{eq:eff_Hk}
	\hH^{(k)} = -\frac{\kappa k}{2}(1-\gamma)-g^2\omega_0\hS_x+(g^2-2)\mri \omega_0\hS_y-2\kappa\gamma\hS_y^2.
\end{equation}
The previously considered $H_1$ and $H_2$ are matrix representations of $\hH^{(1)}$ and $\hH^{(2)}$. The expectation values of all $k$th-order bosonic operators remain finite (are stable), if all eigenvalues of $\hH^{(k)},\hH^{(k-2)},\dotsc$ have negative real parts. Furthermore, the union of the $\hH^{(k)}$ spectra with respect to $k=0,1,2,\dotsc$ yields the full spectrum of the Liouvillian \eqref{eq:mastereqn2}.

The stability boundaries for several even $k$, obtained by exact diagonalization of $\hH^{(k)}$, are shown in Fig.~\ref{fig:boundary}.
As expected, the boundaries exhibit a clear hierarchical structure. As shown in Appx.~\ref{appx:odd-k_up}, the odd-$k$ stability boundaries exhibit an analogous hierarchy, and an instability in $\L_\down$ always occurs before the corresponding instability in $\L_\up$.

Qualitatively, the instability of the bosonic mode arises due to the competition of damping $\mathcal{D}[\ha]$, which shrinks the phase-space extent of the Wigner function, with squeezing $\sim(\ha^2+\ha^{\dag 2})$ due to the Hamiltonian \eqref{eq:Hdown} and dephasing $\mathcal{D}[\hn]$, which cause stretching of the Wigner function in one direction~\cite{gietka2023unique} and angular diffusion, respectively. This gives an intuitive explanation for why $g_c^{(k)}$ is lowered as $\gamma$ increases. Even when the damping strength is fixed, adding small dephasing causes operators of order $k\geq 2$ to diverge at smaller $g$ whenever $\kappa/\omega_0 \neq 2$, which is explained by the interplay of squeezing, rotation, damping, and dephasing of the Wigner function in Appx.~\ref{appx:Wigner}.

For small $\gamma$, we can assess the spectrum of the non-Hermitian Hamiltonians \eqref{eq:eff_Hk} using perturbation theory. To leading order in $\gamma$, the eigenvalue of $\hH^{(k)}$ with the largest real part reads
\begin{equation*}\label{eq:PT_gap}
	\ell^{(k)}_{\text{max}}=k\Big(-\frac{\kappa}{2}+\omega_0\sqrt{g^2-1}+\frac{\kappa\gamma(k-1)(g^2-2)^2}{8(g^2-1)} \Big),
\end{equation*}
as derived in Appx.~\ref{appx:PT}. For the relevant region $g>1$, the real part of $\ell^{(k)}_{\text{max}}$ increases with $k$, which is consistent with the observation that, coming from the normal phase, higher-order observables become unstable before lower-order observables. For $k>1$, the real-parts also increase with $\gamma$, which corresponds to the lowering of the instability thresholds $g_c^{(k)}$.

The $k\to\infty$ limit can be analyzed by treating the spin in Eq.~\eqref{eq:eff_Hk} classically, replacing $\hS_r$ by real numbers $S_r$.
The real part of $H^{(k)}$ is maximized when the spin is fully polarized along the $-x$ direction such that $(S_x,S_y,S_z)=(-k/2,0,0)$. The corresponding $H_k$ value is $-\frac{k}{2}(\kappa-\kappa\gamma-g^2\omega_0)$.
The stability condition requires this value to be negative. Hence, for $\gamma>0$,
\begin{equation}\label{eq:gcinfty}
	g^\infty_c=\sqrt{\kappa(1-\gamma)/\omega_0}
\end{equation}
is the global stability boundary of the model \eqref{eq:mastereqn2} in the thermodynamic limit $\eta \to \infty$, and  Fig.~\ref{fig:boundary} confirms that the $k$th-order boundaries approach $g^\infty_c$ as $k$ increases.

Note that $g^\infty_c < g_c$ except for $(\gamma, \kappa/\omega_0) = (0, 2)$. This leads to two interesting observations:
(a) The normal-phase boundary $g^\infty_c$ is discontinuous at $\gamma=0$ for all $\kappa/\omega_0 \neq 2$. Mathematically, this discontinuity is due to the different scaling of the Gaussian and dephasing components in $\hH^{(k)}$ with respect to $k$. Specifically, the norm of $-2\kappa\gamma\hS_y^2$ grows as $\sim k^3$ while the norm of the remainder grows as $\sim k^2$. This also explains the failure of the perturbative analysis for $k \to \infty$. Physically, the discontinuity implies that dephasing is a relevant perturbation. Arbitrarily small $\gamma$ drives us away from the Gaussian DPT.
(b) For $\gamma>0$ and $g \in (g^\infty_c, g_c)$, the system can be described as an unstable combination of two stable processes: The dephasing dissipator and the remaining Gaussian part of the model are stable individually, but their combination is unstable.
For a different setting, a similar effect is reported in Ref.~\cite{Pocklington2023}, where one of the individually stable components is a Hamiltonian system without dissipation.

\Emph{Non-analyticity in $\expect{\hsigma^z}_s$}
In some experiments, qubit observables are more accessible than oscillator observables. The non-Gaussian DPT can then still be observed through a nonanalytic change in the steady-state expectation value of $\hsigma^z$.
In the $\eta\to\infty$ limit, any convex combination
\begin{equation}
	\dmf_s = p_\down \dmf_{s,\down} + p_\up \dmf_{s,\up}
\end{equation}
of the states \eqref{eq:degenerate} is a steady state, where $p_\down, p_\up \geq 0$ and $p_\down + p_\up = 1$.

For finite $\eta$, the degeneracy is lifted and the steady state $\dmf_s(\eta)$ is unique. We can apply zeroth-order degenerate perturbation theory to find $\lim_{\eta\to\infty}\dmf_s(\eta)$. Its $\hsigma^z$ expectation value takes the form
\begin{subequations}\label{eq:sigmazs}
\begin{equation}
	\lim_{\eta\to\infty}\expect{\hsigma^z}_s= \frac{\gamma(y_\down - y_\up)}{-2(1-\gamma) + \gamma (y_\down + y_\up)}.
\end{equation}
Analytical expressions for
\begin{equation}
	y_\down := \Tr[(\ha-\ha^\dag)^2 \dm_\down], \quad y_\up := \Tr[(\ha-\ha^\dag)^2 \dm_\up]
\end{equation}
\end{subequations}
are derived in Appx.~\ref{appx:sigmaz}.
At $g=g^{(2)}_c$, $|y_\down|$ diverges in the thermodynamic limit while $y_\up$ remains finite as the bosonic $k=2$ operators are still stable with respect to the spin-up Liouvillian $\L_\up$. Thus, we predict the following nonanalytic change in $\expect{\hsigma^z}_s$ at $g^{(2)}_c$.
\begin{equation} \label{eq:sigmazatgc2}
	\text{For all}\ \gamma > 0, \quad
	\lim_{\eta\to\infty}\expect{\hsigma^z}_s \begin{cases}
		< 1 & \text{if}\ g < g^{(2)}_c, \\
		= 1 & \text{if}\ g \geq g^{(2)}_c.
	\end{cases}
\end{equation}
Additionally, Eq.~\eqref{eq:sigmazs} predicts the nonanalytic change in $\expect{\hsigma^z}_s$ at $\gamma=0$ (where $g_c^{(2)}=g_c$).
\begin{equation} \label{eq:sigmazatgamma0}
	\text{For all}\ g \geq g_c, \quad 
	\lim_{\eta\to\infty}\expect{\hsigma^z}_s = \begin{cases}
		0 & \text{if}\ \gamma=0, \\
		1 & \text{if}\ \gamma>0.
    \end{cases}
\end{equation}
Such nonanalytic changes in the qubit polarization are unique to the non-Gaussian DPT of the Rabi model as, in the Gaussian model ($\gamma=0$), $\lim_{\eta\to\infty}\expect{\hsigma^z}_s=0$ is constant. Note that Eqs.~\eqref{eq:sigmazatgc2} and \eqref{eq:sigmazatgamma0} are also in stark contrast to the mean-field analysis which, for all $\gamma$, predicts $\expect{\hsigma^z}_s=-1$ if $g<g_c$ and $\expect{\hsigma^z}_s = -(g_c/g)^2 < 0$ if $g>g_c$; cf.\ Ref.~\cite{hwang2018dissipative} and Appx.~\ref{appx:MFT}. This highlights the aforementioned limitations of mean-field theory.

\Emph{Numerical analysis}
\begin{figure}[t]
	\includegraphics[width=\linewidth]{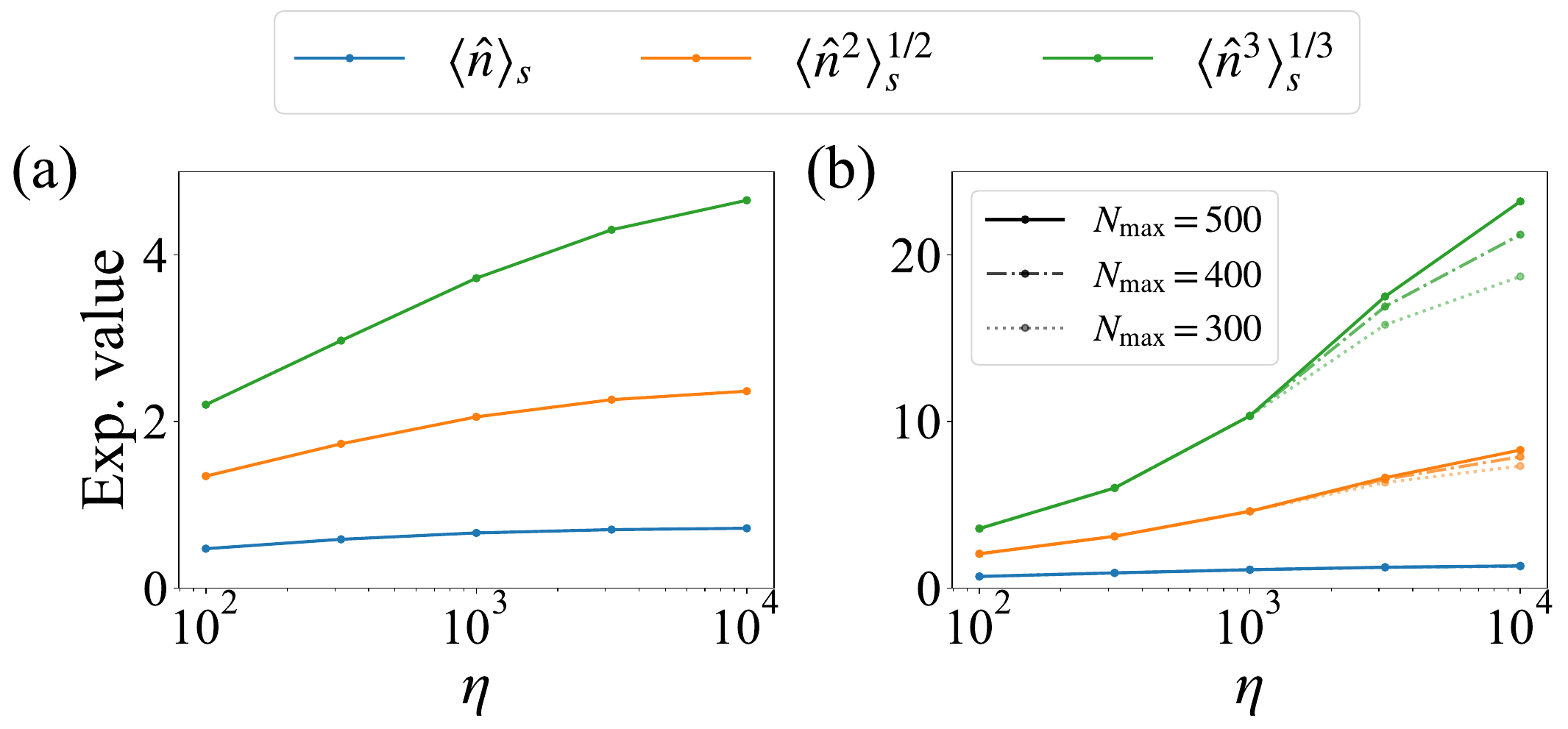}
	\caption{\label{fig:hierarchy} Steady-state expectation values for the 2nd, 4th, and 6th-order operators $\hn$, $\hn^2$, and $\hn^3$ at $\kappa/\omega_0=1$, $g=1.05$ for \textbf{(a)} $\gamma=0.1$ such that $g<g_c^{(6)}$ and \textbf{(b)} $\gamma=0.3$ such that $g_c^{(4)}<g<g_c^{(2)}$. In (a), the expectation values are well converged for $N_{\max}\gtrsim 300$. For (b), we used $N_{\max}=300$, 400, and 500.}
\end{figure}
\begin{figure*}[t]
	\includegraphics[width=0.94\linewidth]{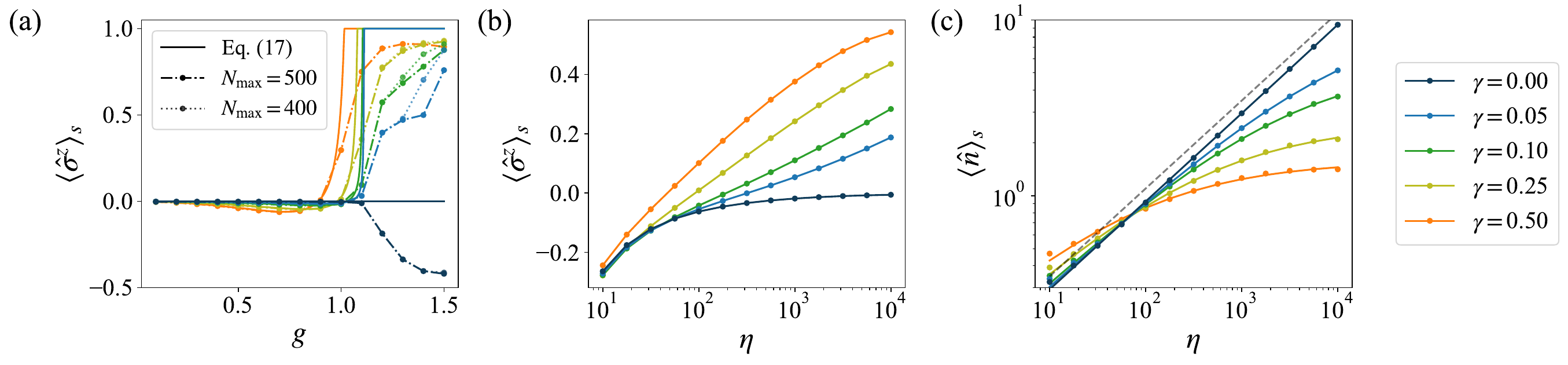}
	\caption{\label{fig:etascaling} \textbf{(a)} Steady-state expectation values of $\hsigma^z$ for various values of $g$ and $\gamma$. Numerical results for $\eta=10^3$ are compared with the $\eta\to\infty$ limit \eqref{eq:sigmazs}. 
	\textbf{(b, c)} Numerically calculated steady-state expectation values of (b) $\hsigma^z$ and (c) $\hn$ for various values of $\eta$ and $\gamma$, where $g=g_c^{(2)}$ for each $\gamma$ value. Expectation values converged well with $N_{\max}=500$. In (c), the dashed line is proportional to $\eta^{1/2}$ and solid lines show fits of the data to the crossover function $\expect{\hn}_s = {B\eta^{1/2}}/{(1+A\eta^{1/2})}$. For all plots, $\kappa/\omega_0=1$.}
\end{figure*}
We numerically evaluate the steady-state expectation values of various observables of the dissipative quantum Rabi model \eqref{eq:mastereqn} with bosonic damping and dephasing. The number of oscillator energy levels is truncated at $N_{\max}$ such that Krylov subspace methods can be employed to the Liouvillian superoperator~\cite{lehoucq1998arpack}.

Figure~\ref{fig:hierarchy} shows the steady-state expectation values of the number operator $\hn$, $\hn^2$, and $\hn^3$. For $(\gamma, g) = (0.1, 1.05)$, all three expectation values grow slowly with $\eta$ before saturating as expected because this point is below the $k=6$ instability boundary. The $\hn^3$ curve has not entirely saturated for the accessible $\eta$, as we are relatively close to $g^{(6)}_c$ (see Fig.~\ref{fig:boundary}). For $(\gamma, g) = (0.3, 1.05)$, only $\expect{\hn}_s$ saturates, and the accessible data is consistent with a divergence of $\expect{\hn^2}_s$ and $\expect{\hn^3}_s$. This is expected from the $\eta\to\infty$ analysis as the point $(0.3, 1.05)$ is between the stability boundaries for $k=2$ and 4. For $\eta>10^3$, the $\expect{\hn^2}_s$ and $\expect{\hn^3}_s$ data is not entirely converged with respect to the cutoff $N_{\max}$ (see also Appx.~\ref{appx:Nmax}). Overall, the data supports the predicted cascade of instabilities, which is a unique feature of the non-Gaussian dissipation with damping and dephasing. Furthermore, the monotonic increase of the expectation values with $\eta$ can be motivated by a small-$\eta$ analysis. As discussed in Appx.~\ref{appx:zeroeta}, small $\eta$ tends to generate a coherent drive term that is proportional to $\ha+\ha^\dag$, while large $\eta$ generates a squeezing term [included in $(\ha+\ha^\dag)^2$; see Eq.~\eqref{eq:Hdown}] instead, where only the latter accounts for the instability.

For various values of $g$ and $\gamma$, Fig.~\ref{fig:etascaling}a compares the numerically evaluated $\expect{\hsigma^z}_s$ at $\eta=10^3$ to the $\eta\to\infty$ prediction~\eqref{eq:sigmazs}. For all $\gamma$, the numerical results agree well with the analytical expression in the $g < g_c^{(2)}$ regime. For $\gamma>0$ and $g$ being increased beyond $g_c^{(2)}$, the numerical $\expect{\hsigma^z}_s$ data rapidly increases to a value near 1 in accordance with the predicted nonanalytic change \eqref{eq:sigmazatgc2} for $\eta\to\infty$.

Figure~\ref{fig:etascaling}b shows the $\eta$ dependence of $\expect{\hsigma^z}_s$ for different $\gamma$ at $g=g_c^{(2)}$, i.e., points along the $k=2$ instability boundary in Fig.~\ref{fig:boundary}. For $\gamma>0$, $\expect{\hsigma^z}_s$ increases monotonically and has not yet saturated at large $\eta$, suggesting that the deviations between the numerical data and Eq.~\eqref{eq:sigmazs} observed in Fig.~\ref{fig:etascaling}a are due to finite-$\eta$ effects. For $\gamma=0$, $\expect{\hsigma^z}_s$ is seen to converge to 0 such that, overall, the observations are consistent with the prediction~\eqref{eq:sigmazatgamma0}.

For the Gaussian system ($\gamma=0$) at the transition point $g=g_c$, the oscillator occupation scales according to the power law $\expect{\hn}_s \propto \eta^{1/2}$ as shown using Keldysh field theory in Ref.~\cite{hwang2018dissipative} and Appx.~\ref{appx:Keldysh}. Figure~\ref{fig:etascaling}c confirms this scaling numerically but also shows that, for the non-Gaussian $\gamma>0$, $\expect{\hn}_s$ crosses over to a constant scaling at larger $\eta$, substantiating further that dephasing is a relevant perturbation, leading to a non-Gaussian DPT.

\Emph{Discussion and outlook}
We have found that dephasing of the bosonic mode is a relevant perturbation in the open quantum Rabi model. In the thermodynamic limit $\eta\to\infty$, infinitesimal dephasing induces nonanalytic changes in both the normal-phase boundary and the qubit polarization, fundamentally changing the physics compared to prior work \cite{hwang2015quantum}.
The bosonic observables display an unusual cascade of instabilities, where higher-order observables diverge before lower-order ones. The change in $\expect{\hsigma^z}_s$ with respect to the non-Gaussianity parameter $\gamma$ is of particular interest, as Eq.~\eqref{eq:dampdephsame} implies that both the mean-field analysis and the leading-order characterization of the bosonic mode in terms of its autocorrelation functions and spectral density~\cite{tamascelli2018nonperturbative, sun2024quantum} are insensitive to $\gamma$. 

The tree-level scaling analysis of the model's Keldysh action in Appx.~\ref{appx:Keldysh} confirms that both dephasing and finite-$\eta$ corrections are renormalization-group relevant perturbations that drive the system away from the Gaussian critical point. A further field-theoretical analysis could reveal the universal properties of the resulting non-Gaussian DPT at $g=g_c^\infty$. 
We anticipate that non-Gaussian effects give rise to distinctive phenomena in other open systems like the Dicke model \cite{hepp1973superradiant, wang1973phase, emary2003chaos, nagy2011critical, torre2013keldysh, kirton2017suppressing, gegg2018superradiant} and Leggett's spin-boson model \cite{Leggett87, nalbach2013crossover, strathearn2018efficient}. Further non-Gaussian effects to be considered are, for example, oscillator anharmonicities and non-linear qubit-oscillator couplings~\cite{Kang24,mcdonald2022exact}. 

Another important contribution of this Letter is the connection between the Green's function EOM of the open system and simple non-Hermitian spin Hamiltonians. This connection makes it possible to efficiently determine the spectrum of the quartic Liouvillians through exact diagonalization, perturbation theory, and semi-classical limits. As the condition for the closed hierarchy only requires the bilinear Lindblad operators to be Hermitian, this trick can be applied for large classes of Markovian quantum dynamics, establishing a new bridge between driven-dissipative systems and non-Hermitian quantum mechanics~\cite{Bender2007-70,Moiseyev2011,Ashida2020}.

Experimentally, the non-Gaussian DPT and instability cascade in the quantum Rabi model could, for example, be investigated using trapped ions~\cite{cai2021observation} and circuit quantum electrodynamics~\cite{zheng2023observation}.
For trapped ions, tunable damping and dephasing of the motional mode have recently been implemented~\cite{pagano2025fundamentals} using sympathetic cooling~\cite{so2024trapped} and stochastic modulations of driving fields~\cite{sun2024quantum}, respectively. Similarly, decoherence processes can be engineered and tuned in circuit quantum electrodynamics~\cite{harrington2022engineered, maurya2024demand}. In general, quantum simulators are promising tools for the study of open quantum systems beyond Gaussian models~\cite{Kang24,macdonell2021analog,sun2024quantum}.

\begin{acknowledgments}
We gratefully acknowledge helpful feedback by Yu-Xin Wang and support from the U.S.\ National Science Foundation (grant PHY-2412555), the Quantum Leap Challenge Institute for Robust Quantum Simulation (grant OMA-2120757), and the U.S.\ Army Research Office (MURI grant W911NF1810218).
\end{acknowledgments}\vspace{1ex}

\Emph{Data Availability} The data that support the findings of this article are openly available \cite{Kang2025_07data}.

\newpage

\appendix

\onecolumngrid
\section{Mean-field analysis}\label{appx:MFT}
The evolution of observables can be deduced from the adjoint Lindblad master equation such that
\begin{equation}\label{eq:LMEadj}
	\partial_t\expect{\hA}=\expect{\L^\dag(\hA)}\qq{with}
	\L^\dag(\hA)
	= \mri[\hH_\text{Rabi}, \hA] + \kappa(1-\gamma)\D^\dag[\ha](\hA) + \kappa\gamma \D^\dag[\hn](\hA).
\end{equation}
For the bosonic annihilation operator $\ha$ and the Pauli operators, we have
\begin{subequations}\label{eq:MFT-EOM}
\begin{align}\label{eq:MFT-a}
	\L^\dag(\ha)&=-\Big(\frac{\kappa}{2}+\mri\omega_0\Big)\,\ha-\mri\lambda\hsigma^x,\\ \label{eq:MFT-x}
	\L^\dag(\hsigma^x)&=-\Omega\hsigma^y,\\ \label{eq:MFT-y}
	\L^\dag(\hsigma^y)&=\Omega\hsigma^x-2\lambda\hsigma^z(\ha+\ha^\dag),\qq{and}\\ \label{eq:MFT-z}
	\L^\dag(\hsigma^z)&=2\lambda\hsigma^y(\ha+\ha^\dag).
\end{align}
\end{subequations}

With the mean-field variables
\begin{equation}
	a:=\expect{\ha}/\sqrt{\eta},\quad
	s_r:=\expect{\hsigma^r}\qq{and approximation}
	\expect{\hsigma^r\ha}\approx\expect{\hsigma^r}\expect{\ha},
\end{equation}
the steady-state equations $\partial_t\expect{\ha}=0$, $\partial_t\expect{\hsigma^r}=0$ imply $s_y=0$ due to Eq.~\eqref{eq:MFT-x},
\begin{subequations}\label{eq:MFT-ss}
\begin{alignat}{6}
	0&=\partial_t\expect{\ha} &&\stackrel{\eqref{eq:MFT-a}}{=}
	-\Big(\frac{\kappa}{2}+\mri\omega_0\Big)\,\sqrt{\eta}\,a-\mri\lambda s_x\quad
	&&\Leftrightarrow\quad 
	0 = \Big(1-\mri\frac{\kappa}{2 \omega_0}\Big)\,a+\frac{g}{2}\,s_x,\qq{and}\\
	0&=\partial_t\expect{\hsigma^y} &&\stackrel{\eqref{eq:MFT-y}}{=}
	\Omega s_x-2\lambda s_z\sqrt{\eta}\,(a+a^*)\quad
	&&\Leftrightarrow\quad
	0 = s_x - g\,(a+a^*)\,s_z.
\end{alignat}
\end{subequations}

As pointed out in the  main text, oscillator damping $\propto \kappa(1-\gamma)$ and dephasing $\propto\kappa\gamma$ have the same effect on the evolution of $\expect{\ha}$. Hence, the mean-field equations depend on $\kappa$ but are independent of the non-Gaussianity parameter $\gamma$. Consequently, the mean-field steady-state solutions agree with those of the dephasing-free Rabi model discussed in Ref.~\cite{hwang2018dissipative}. Specifically, in conjunction with the constraint $s_x^2+s_y^2+s_z^2=1$, the mean-field steady-state equations ~\eqref{eq:MFT-ss} have the stable solutions
\begin{subequations}
\begin{alignat}{4}
	&a=s_x=s_y=0,\quad s_z=\pm1
	 &&\qq{for}g\leq g_c\qq{and}\\
	&a=\pm\frac{1}{2}\,\frac{g}{1-\mri\kappa/(2\omega_0)}\sqrt{1-\left(\frac{g_c}{g}\right)^4},\ \
	s_x=\mp \sqrt{1-\left(\frac{g_c}{g}\right)^4},\ \
	s_y=0,\ \, s_z=-\left(\frac{g_c}{g}\right)^2
	 &&\qq{for}g>g_c.
\end{alignat}
\end{subequations}
So, the mean-field theory suggests a DPT at $g=g_c=\sqrt{1+\kappa^2/(4\omega_0^2)}$ with spontaneous breaking of the weak $\mathbb{Z}_2$ symmetry and critical exponent $\beta=1/2$ for the order parameter $a=\expect{\ha}/\sqrt{\eta}$.

\section{The \texorpdfstring{$k=2$}{k=2} stability boundary for \texorpdfstring{$\L_\down$}{Ldown}}\label{appx:k2}
For the soft-mode limit $\eta \to \infty$ and the qubit being in the $\ket{\down}$ state, we show in the following that the second-order bosonic operators are stable if and only if
\begin{equation}\label{eq:gc2app}
	g < g^{(2)}_c := \left(\frac{\sqrt{1-\gamma}}{\gamma}
	 \left(\sqrt{1+\gamma+\frac{\gamma\kappa^2}{2\omega_0^2}(1+\gamma)^2} - \sqrt{1-\gamma}\right) \right)^{1/2}.
\end{equation}

From the adjoint Lindblad master equation with $(\L,\hH_\text{Rabi})\to(\L_\down,\hH_\down)$ in Eq.~\eqref{eq:LMEadj}, we can write the EOM for the expectation-value vector $\vec{v}_2 := (\expect{\ha\ha}, \expect{\ha^\dag\ha}, \expect{\ha^\dag\ha^\dag})^T$ of the second-order bosonic operators as
\begin{equation} \label{eq:k2EOM}
	\partial_t\vec{v}_2 = H_2\vec{v}_2+Y_2
	= \mri\omega_0 \begin{pmatrix}
		\mri\frac{(1+\gamma)\kappa}{\omega_0} - 2+g^2 & g^2 & 0 \\
		-\frac{g^2}{2} & \mri\frac{(1-\gamma)\kappa}{\omega_0} & \frac{g^2}{2} \\
		0 & -g^2 & \mri\frac{(1+\gamma)\kappa}{\omega_0} + 2-g^2
	\end{pmatrix} \vec{v}_2
	+ \mri\omega_0 \begin{pmatrix} \frac{g^2}{2} \\ 0 \\ -\frac{g^2}{2} \end{pmatrix}.
\end{equation}
This system of first-order differential equations is stable if and only if all eigenvalues of $H_2$ have negative real parts. This can be checked by the Routh-Hurwitz criterion~\cite{hurwitz1964conditions}. The characteristic polynomial of $H_2$ is
\begin{equation}
	P(\ell) = \det(\ell\id - H_2) = \ell^3 + a_2 \ell^2 + a_1 \ell + a_0
\end{equation}
where
\begin{subequations}\label{eq:afactor}
\begin{align}
	a_2 &= (3+\gamma) \kappa, \\
	a_1 &= (3-\gamma)(1+\gamma)\kappa^2 - 4(g^2-1)\omega_0^2,\qq{and} \\
	a_0 &= (1-\gamma)(1+\gamma)^2 \kappa^3 - 4\Big(\frac{\gamma g^4}{2} + (1-\gamma)(g^2-1) \Big) \kappa\omega_0^2.
\end{align}
\end{subequations}

According to the Routh-Hurwitz criterion for monic third-order polynomials \cite{hurwitz1964conditions}, all eigenvalues of $H_2$ have negative real parts if and only if
\begin{equation}\label{eq:RH}
	a_0,a_1,a_2 > 0\qq{and} a_2a_1 - a_0 > 0.
\end{equation}

\noindent\textbf{(a)} As $\kappa>0$ and $\gamma\geq 0$, we have $a_2 > 0$ for all $g$.

\noindent\textbf{(b)} The condition $a_1 > 0$ holds for all $g < g^{(1)}_c$ as
\begin{align}\nonumber
	a_1 &= (3-\gamma)(1+\gamma)\kappa^2 - 4(g^2-1)\omega_0^2 \\\nonumber
	&>  (3-\gamma)(1+\gamma)\kappa^2 - \kappa^2 
	= (2+2\gamma-\gamma^2) \kappa^2\\
	&\geq 2 \kappa^2 > 0.
\end{align}
where, the first inequality follows from $g^2-1 < (g^{(1)}_c)^2-1 = \kappa^2/4\omega_0^2$, and the second inequality follows from $0\leq \gamma\leq 1$.

\noindent\textbf{(c)} Next, we prove that $a_0 > 0$ if and only if $g < g^{(2)}_c$. Consider $a_0$ as a function of $h := g^2$ such that
\begin{equation}\label{eq:appx_a0}
	\frac{a_0(h)}{\kappa\omega_0^2} = -2\gamma h^2 - 4(1-\gamma)h + (1-\gamma)\left( (1+\gamma)^2 \frac{\kappa^2}{\omega_0^2} + 4 \right).
\end{equation}
Note that $a_0(0) \geq 0$ and $\lim_{h\to\infty}a_0(h)= -\infty$. Thus, as $a_0$ is a quadratic function of $h$, $a_0 > 0$ if and only if $h < h'$ where $a_0(h') = 0$. Straightforward algebra yields $h' = (g^{(2)}_c)^2$.

\noindent\textbf{(d)} We can show that $a_2a_1-a_0>0$ if $g < g^{(1)}_c$ as follows: 
\begin{align*}
    \frac{a_2a_1 - a_0}{\kappa\omega_0^2} &= (3+\gamma) \Big( 
    (3-\gamma)(1+\gamma)\frac{\kappa^2}{\omega_0^2} - 4(g^2-1)
    \Big) - (1-\gamma)(1+\gamma)^2 \frac{\kappa^2}{\omega_0^2} + 4\Big(\frac{\gamma g^4}{2} + (1-\gamma)(g^2-1) \Big) \\
    &= 8(1+\gamma)\frac{\kappa^2}{\omega_0^2} - 8(g^2-1) + 2\gamma (g^2-2)^2 \\
    &= 32(1+\gamma)\big((g^{(1)}_c)^2-1\big) - 8(g^2-1) + 2\gamma (g^2-2)^2 \\
    &> 32(1+\gamma)\big((g^{(1)}_c)^2-1\big) - 8\big((g^{(1)}_c)^2-1\big)\\
    &= 8(3+4\gamma)\big((g^{(1)}_c)^2-1\big) > 0.
\end{align*}

\noindent\textbf{(e)} Finally, straightforward calculations show that $\lim_{\gamma\to 0}g^{(2)}_c = g^{(1)}_c$, and $g^{(2)}_c$ is a decreasing function of $\gamma\in[0,1]$ while $g^{(1)}_c$ is constant. Thus, $g^{(2)}_c \leq g^{(1)}_c$. This completes the proof that the Routh-Hurwitz criterion is satisfied if $g < g^{(2)}_c$. Also, according to point (c), $g < g^{(2)}_c$ only if the Routh-Hurwitz criterion is satisfied.

\section{Mapping the EOM for \texorpdfstring{$k$}{k}-point Green’s functions to a non-Hermitian spin-\texorpdfstring{$k/2$}{k/2} Hamiltonian}\label{appx:eff_H}
For the soft-mode limit $\eta \to \infty$ and the qubit being in the $\ket{\down}$ state, we will see in the following how the EOM for the bosonic $k$th-order observables can be mapped to an effective non-Hermitian Hamiltonian for a spin-$k/2$ model.

Following the convention in Ref.~\cite{Barthel2022}, we first perform a unitary transformation from the bosonic ladder operators $\ha$ and $\ha^\dag$ to 
\begin{equation}\label{eq:majorana}
    \begin{pmatrix} \hw_+ \\ \hw_- \end{pmatrix}
    =\frac{1}{\sqrt{2}}\begin{pmatrix} 1&1 \\ \mri&-\mri \end{pmatrix}
      \begin{pmatrix} \ha \\ \ha^\dag \end{pmatrix},
    \qq{ such that}
    [\hw_+,\hw_-]=-\mri.
\end{equation}
The self-adjoint operators $\hw_\pm$ are, up to a sign factor, the oscillator's position and momentum operators.
With these, the Hamiltonian \eqref{eq:Hdown} can be written in the form
\begin{equation}
    \hH_\down=\begin{pmatrix}\hw_+, \hw_- \end{pmatrix} H_\down \begin{pmatrix}\hw_+ \\ \hw_-\end{pmatrix},
    \qq{with} H_\down =\frac{\omega_0}{2}\begin{pmatrix}1-g^2 &0 \\ 0& 1 \end{pmatrix}.
\end{equation}
The linear Lindblad operator $\hL$ for damping and the quadratic Lindblad operator $\hM$ for dephasing take the form
\begin{align}
	\hL&=\sqrt{\kappa(1-\gamma)/2}\,\ha=\sqrt{\kappa(1-\gamma)/2}\,(\hw_+-\mri\hw_-)\qq{and}\\
	\hM&=\sqrt{\kappa\gamma}\,\ha^\dag\ha
	=\begin{pmatrix}\hw_+, \hw_-\end{pmatrix} M\begin{pmatrix} \hw_+ \\ \hw_-\end{pmatrix}
	\qq{with} M=\frac{\sqrt{\kappa\gamma}}{2}\id.
\end{align}
As in Ref.~\cite{Barthel2022}, we also define the following $2\times 2$ matrices for later convenience.
\begin{subequations}\label{eq:BXYZ}
\begin{align}
    B&:=\frac{\kappa(1-\gamma)}{2}\begin{pmatrix}
    1&\mri \\ -\mri&1
    \end{pmatrix},
    \quad
    B_r:=\frac{B+B^\dag}{2}=\frac{\kappa(1-\gamma)}{2}\id,
    \quad B_i:=\frac{B-B^\dag}{2\mri}=\frac{\kappa(1-\gamma)}{2}\mri\sigma^y, \\
    X&:=-2\mri\sigma^y H_\down +\mri\sigma^yB_i-2(\sigma^y M)^2
       =-\frac{\kappa}{2}\id + \omega_0\begin{pmatrix}0&-1 \\ 1-g^2&0\end{pmatrix}, \\
    Y&:=\sigma^y B_r\sigma^y=\frac{\kappa(1-\gamma)}{2}\id,\qq{and}
    Z:=2\mri\sigma^yM=\sqrt{\kappa\gamma}\,\mri\sigma^y.
\end{align}
\end{subequations}

The bosonic $k$th-order observables are fully characterized by the symmetrized static $k$-point Green's function
\begin{equation}\label{eq:Gk-def}
	\Gamma^{(k)}_{\mu_1\cdots\mu_k}:=\frac{1}{k!}\sum_{\sigma}\expect{ \hw_{\mu_{\sigma(1)}}\cdots\hw_{\mu_{\sigma(k)}} }
\end{equation}
with $\mu_i\in\{+,-\}$ and the summation running over all permutations $\sigma$ of $k$ objects. $\Gamma^{(k)}\in (\mathbb{R}^2)^{\otimes k}$ is a real symmetric tensor of order $k$ with dimensions $(2,2,\dotsc,2)$, similar to a permutation-symmetric pure quantum state for $k$ spins-1/2 (qubits).
For example,
\begin{equation}
    \Gamma^{(2)}=\begin{pmatrix}
    \expect{\hw_{+}^2}& (\expect{\hw_+\hw_-}+\expect{\hw_-\hw_+})/2 \\
    (\expect{\hw_+\hw_-}+\expect{\hw_-\hw_+})/2& \expect{\hw_{-}^2}
    \end{pmatrix}.
\end{equation}
The symmetrization in Eq.~\eqref{eq:Gk-def} does not lead to any loss of information because, according to the commutation relation \eqref{eq:majorana}, the expectation values $\expect{ \hw_{\mu_{\sigma(1)}}\cdots\hw_{\mu_{\sigma(k)}} }$ for different permutations $\sigma$ of the indices $\mu_i$ agree up to a constant and lower-order Green's function terms.

Using the EOM
\begin{equation}\label{eq:LMEadj2}
	\partial_t\expect{\hA}=\expect{\L_\down^\dag(\hA)}\qq{with}
	\L_\down^\dag(\hA)=\mri [\hH_\down,\hA]+\frac{1}{2}\hL^\dag[\hA,\hL]+\frac{1}{2}[\hL^\dag,\hA]\hL-\frac{1}{2}[\hM,[\hM,\hA]],
\end{equation}
for the expectation values, the commutation relation \eqref{eq:majorana}, and the commutator identity
\begin{equation}
	[\hA,\hB_1\hB_2\cdots \hB_k]\ =\ [\hA,\hB_1]\hB_2\cdots \hB_k\,+\,\hB_1[\hA,\hB_2]\hB_3\cdots \hB_k\,+\,\dotsc\,+\,\hB_1\cdots\hB_{k-1}[\hA,\hB_k],
\end{equation}
we arrive at the following EOM for $\Gamma^{(k)}$:
\begin{equation}\label{eq:EOM-Gk}
    \partial_t\Gamma^{(k)}_{\mu_1\cdots\mu_k}=
    \sum_{i=1}^k\sum_{\alpha}X_{\mu_i\alpha}\underbrace{\Gamma^{(k)}_{\mu_1\cdots\alpha\cdots\mu_k}}_{\text{replace $\mu_i$ by $\alpha$}}
    +\frac{1}{2}\sum_{i\neq j}Y_{\mu_i\mu_j}\underbrace{\Gamma^{(k-2)}_{\mu_1\cdots\mu_k}}_{\text{remove $\mu_i, \mu_j$}}
    +\frac{1}{2}\sum_{i\neq j}\sum_{\alpha,\beta}Z_{\mu_i \alpha}Z_{\mu_j \beta}\underbrace{\Gamma^{(k)}_{\mu_1\cdots\alpha\cdots\beta\cdots\mu_k}}_{\text{replace $\mu_i,\mu_j$ by $\alpha,\beta$}}\!\!\!,
\end{equation}
with the $2\times2$ matrices $X,Y,Z$ from Eq.~\eqref{eq:BXYZ} and $\alpha,\beta\in\{+,-\}$.

The EOM \eqref{eq:EOM-Gk} shows that $\partial_t\Gamma^{(k)}$ only depends on $\Gamma^{(k)}$ and $\Gamma^{(k-2)}$. In other words, the EOM for $(\Gamma^{(1)},\Gamma^{(2)},\cdots)^T$ assume a block-triangular form. After vectorization of $\Gamma^{(k)}$, the corresponding $k$-th diagonal block of the generator of the time evolution can be written in the form
\begin{equation}\label{eq:eff_Hk_collective}
	\hH^{(k)}=-\frac{\kappa k}{2}\id+\sum_{i=1}^k(\underbrace{-\omega_0\hsigma_i^+ + (1-g^2)\omega_0\hsigma_i^-}_{=-g^2\omega_0\hsigma_i^x/2+(g^2-2)\mri\omega_0\hsigma_i^y/2})-\frac{\kappa\gamma}{2}\sum_{i\neq j}\hsigma_i^y\hsigma_j^y,
\end{equation}
which can be interpreted as a non-Hermitian Hamiltonian for a $k$-site spin-$1/2$ system, where $\hsigma_i^r$ are the Pauli operators for the $i$-th spin.
The spin model \eqref{eq:eff_Hk_collective} is permutation symmetric, i.e., a so-called \emph{collective} model. Defining the total spin operators
\begin{equation}
    \hS_x:=\frac{1}{2}\sum_{i=1}^k\hsigma_i^x,\quad
    \hS_y:=\frac{1}{2}\sum_{i=1}^k\hsigma_i^y,\quad
    \hS_z:=\frac{1}{2}\sum_{i=1}^k\hsigma_i^z,
\end{equation}
the model assumes the simple form
\begin{equation}\label{eq:eff_Hk_appx}
	\hH^{(k)}=-\frac{\kappa k}{2}(1-\gamma)-g^2\omega_0\hS_x+(g^2-2)\mri \omega_0\hS_y-2\kappa\gamma\hS_y^2.
\end{equation}
As $[\hS^2,\hS_x]=0$ and $[\hS^2,\hS_y]=0$, the spin magnitude $\hS^2$ is conserved.

Due to the symmetrization, the tensors $\Gamma^{(k)}$ as defined in \eqref{eq:Gk-def} carry redundant information: Interpreting $\mu_i=\pm \mapsto \pm \frac{1}{2}$ as the single-spin $z$ magnetizations, all tensor elements with fixed total $z$ magnetization $M_z$ are the same such that we can define the reduced Green's functions
\begin{equation}\label{eq:Gk-def2}
    \tilde{\Gamma}^{(k)}_{M_z}:=\Gamma^{(k)}_{\mu_1\cdots\mu_k}\qq{with}
	 M_z=\sum_i\mu_i\in\left\{-\frac{k}{2},-\frac{k}{2}+1,\dotsc,\frac{k}{2}\right\},
\end{equation}
which only contain independent expectation values -- one for each of the $k+1$ independent $k$th-order operators
\begin{equation}\label{eq:kthOrderOp}
	\{(\ha^\dag)^m \ha^{k-m}\}\qq{or, equivalently,}
	\{\hw_+^m \hw_-^{k-m}\}\qq{with} m=0,1,\dotsc,k.
\end{equation}
Recall that the ladder operators $\ha$, $\ha^\dag$ are related to $\hw_\pm$ through the unitary transformation \eqref{eq:majorana}.
In conjunction with the $\hS^2$ conservation, we can hence restrict the model \eqref{eq:eff_Hk_appx} to the maximum $\hS^2$ eigenspace, which is $(k+1)$-dimensional, containing exactly one $\hS_z$ eigenstate for each magnetization $M_z$. In other words, the generator for the evolution of the reduced Green's functions \eqref{eq:Gk-def2} is block-triangular, with the $k$-th diagonal block given by the non-Hermitian spin-$k/2$ Hamiltonian \eqref{eq:eff_Hk_appx}.
As the sets \eqref{eq:kthOrderOp} form complete operator bases, the union of the $\hH^{(k)}$ spectra over $k=0,1,\dotsc$ gives the full spectrum of the Liouvillian $\L_\down$. This approach allows for an efficient determination of the stability boundaries $g_c^{(k)}$.

$H_1$ from Eq.~\eqref{eq:M1} is a matrix representation of $\hH^{(1)}$, and $H_2$ in Eq.~\eqref{eq:k2EOM} is a matrix representation of $\hH^{(2)}$.

\section{Stability analysis for odd operator orders and spin-up steady states}\label{appx:odd-k_up}
In the main text, we have discussed the cascade of instabilities for even operator orders $k$ and the qubit being in the spin-$\down$ state, i.e., $\L_\down$ from Eq.~\eqref{eq:mastereqn2}. Specifically, for coupling strengths $g>g^{(k)}_c$ the expectation values of $k$th-order operators $(\ha^\dag)^m \ha^{k-m}$ are unstable in the sense that some of them diverge in the thermodynamic limit $\eta\to\infty$ but they are all stable for $g<g_c^{(k)}$ with $g_c^{(k+2)}\leq g_c^{(k)}$, i.e., higher-order observables diverge first.
In the following, we also consider odd $k$ and the case in which the qubit is in the spin-$\up$ state.
\begin{figure}[t]
\includegraphics[width=\textwidth]{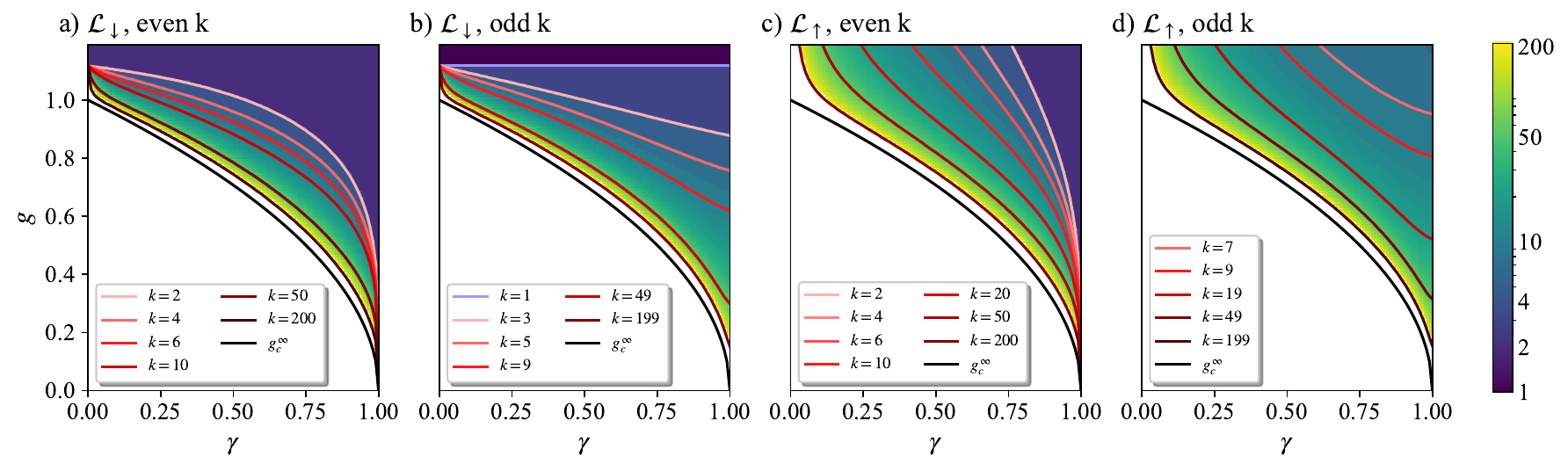}
	\caption{\label{fig:boundary2} Stability boundaries of the $k$th-order bosonic operators in the of thermodynamic limit $\eta \to \infty$ for $\L_\down$ in panels \textbf{(a)} and \textbf{(b)} as well as $\L_\up$ in panels \textbf{(c)} and \textbf{(d)}.
	As in Fig.~\ref{fig:boundary}, which is reproduced in panel (a), the color map always indicates the smallest even or odd $k$ for which, at the given point in the phase diagram, the corresponding non-Hermitian Hamiltonian $\hH^{(k)}$, governing the Green's function evolution, is unstable. For this data, we have set $\kappa/\omega_0 = 1$. }
\end{figure}

Figure~\ref{fig:boundary2} shows a numerical analysis of the corresponding non-Hermitian Hamiltonians $\hH^{(k)}_\down$ as given by Eq.~\eqref{eq:eff_Hk_appx} [Eq.~\eqref{eq:eff_Hk}] and $\hH^{(k)}_\up$, which is given by Eq.~\eqref{eq:eff_Hk_appx} with the substitution $g^2\to -g^2$. Indeed, for spin-$\up$ as well as odd $k$, we obtain similar hierarchies of the stability boundaries $g^{(k)}_{c,\down}$ and $g^{(k)}_{c,\down}$, where all four sets (spin up or down, even or odd $k$) converge to the same global stability boundary $g^{\infty}_c$ at large $k$ as given by Eq.~\eqref{eq:gcinfty}.

The analysis also confirms that
\begin{equation}\label{eq:downdecidesstab}
	g^{(k)}_c\equiv g^{(k)}_{c,\down}\leq  g^{(k)}_{c,\up}\quad\text{for all}\ k,
\end{equation}
i.e., the spin-$\down$ sector always decides the stability.

For $k=1$, and $k=2$, we can assess the stability of
\begin{equation}\label{eq:mastereqn2-up}
    \L_\up=-\mri\Big[\frac{\omega_0g^2}{4}(\ha+\ha^\dag)^2+\omega_0\hn\,,\,\,\cdot\,\,\Big]+\kappa(1-\gamma)\D[\ha]+\kappa\gamma\D[\hn]
\end{equation}
following the analysis of $\L_\down$ in the main text and Appx.~\ref{appx:k2}.
With $g^2\to-g^2$, the corresponding dynamic matrices $H_{1,\up}$ and $H_{2,\up}$ read
\begin{equation}
    H_{1,\up} = \mri\omega_0 \begin{pmatrix}
        \mri \frac{\kappa}{2\omega_0} - 1-\frac{g^2}{2} & -\frac{g^2}{2} \\
        \frac{g^2}{2} & \mri\frac{\kappa}{2\omega_0}+ 1+\frac{g^2}{2}
    \end{pmatrix},\quad
    H_{2,\up} = \mri\omega_0 \begin{pmatrix}
        \mri\frac{(1+\gamma)\kappa}{\omega_0} - 2-g^2 & -g^2 & 0 \\
        \frac{g^2}{2} & \mri\frac{(1-\gamma)\kappa}{\omega_0} & -\frac{g^2}{2} \\
        0 & g^2 & \mri\frac{(1+\gamma)\kappa}{\omega_0} + 2+g^2
    \end{pmatrix}.
\end{equation}

The $H_{1,\up}$ eigenvalues are $-\kappa/2\pm\mri\omega_0\sqrt{1+g^2}$. For all $g$, they have negative real parts, such that the $k=1$ block is always stable, i.e., $g^{(1)}_{c,\up}=\infty$.

For $H_{2,\up}$, we follow the derivation in Appx.~\ref{appx:k2}. Similar to Eq.~\eqref{eq:afactor}, in this case, we have
\begin{subequations}
\begin{align}
    a_{2\up} &= (3+\gamma) \kappa, \\
    a_{1\up} &= (3-\gamma)(1+\gamma)\kappa^2 + 4(g^2+1)\omega_0^2,\qq{and} \\
    a_{0\up} &= (1-\gamma)(1+\gamma)^2 \kappa^3 - 4\Big(\frac{\gamma g^4}{2} - (1-\gamma)(g^2+1) \Big) \kappa\omega_0^2.
\end{align}
\end{subequations}
We immediately realize that $a_{1\up}>0$ and $a_{2\up}>0$. We also have
\begin{align*}
    \frac{a_{2\up}a_{1\up} - a_{0\up}}{\kappa\omega_0^2} = 8(1+\gamma)\frac{\kappa^2}{\omega_0^2} + 8(g^2+1) + 2\gamma (g^2-2)^2 >0.
\end{align*}
So, according to the Routh-Hurwitz criterion \eqref{eq:RH}, $H_{2,\up}$ is stable if and only if
\begin{equation*}
    a_{0\up} = (1-\gamma)(1+\gamma)^2 \kappa^3 - 4\Big(\frac{\gamma g^4}{2} - (1-\gamma)(g^2+1) \Big) \kappa\omega_0^2>0.
\end{equation*}
This leads to the stability boundary
\begin{equation}\label{eq:gc2up}
	g^{(2)}_{c,\up} := \left(\frac{\sqrt{1-\gamma}}{\gamma}
    \left(\sqrt{1+\gamma+\frac{\gamma\kappa^2}{2\omega_0^2}(1+\gamma)^2} + \sqrt{1-\gamma}\right) \right)^{1/2}
\end{equation}
of the $k=2$ block.
The comparison with Eq.~\eqref{eq:gc2app} [Eq.~\eqref{eq:gc2}] shows that $g_c^{(2)} \leq g^{(2)}_{c,\up}$ with equality holding only at $\gamma=1$. Also note that $\lim_{\gamma\to 0}g_c^{(2)}= g_c$ but $\lim_{\gamma\to 0}g^{(2)}_{c,\up}= \infty$.

Indeed, for any finite $k$, $\lim_{\gamma\to 0} g_c^{(k)}= g_c$ and $\lim_{\gamma\to 0} g_{c,\up}^{(k)}=\infty$ as illustrated by the numerical analysis in Fig.~\ref{fig:boundary2}. This is intriguing as both $g_c^{(k)}$ and $g_{c,\up}^{(k)}$ approach $g_c^\infty$ in the $k\to\infty$ limit, and $\lim_{\gamma\to 0}g_c^\infty=1$.

\section{Wigner function analysis} \label{appx:Wigner}
This appendix develops an intuitive understanding for some of the observed phenomena assessing the interplay of squeezing, rotation, damping, and dephasing induced by the Liouvillians $\L_\down$ [Eq.~\eqref{eq:mastereqn2}] and $\L_\up$ [Eq.~\eqref{eq:mastereqn2-up}] for the $\eta\to\infty$ limit in terms of Wigner functions
\begin{equation}
    W(x,p) := \frac{1}{2\pi\hbar} \int_{-\infty}^\infty\ud y\,  e^{ipy/\hbar}\,\bra{x - \frac{y}{2} }\hat{\rho} \ket{ x + \frac{y}{2}},
\end{equation}
where we use natural units with $\hbar=1$ in the following.

This can qualitatively explain the following features of the non-Gaussian DPT: (i) as $g$ is increased from below, an instability in $\L_\down$ occurs before the corresponding instability in $\L_\up$ [Eq.~\eqref{eq:downdecidesstab}], (ii) dephasing has a heating effect and immediately induces instabilities if there is no damping ($\gamma=1$), and (iii) near the Gaussian critical point $(\gamma, g) = (0, g_c)$, adding small dephasing causes the $k$th-order operators $(k\geq 2)$ to diverge at a smaller $g$ value, except when $\kappa/\omega_0 = 2$ [$g_c=g_c^\infty(\gamma=0)=\sqrt{2}$]. 

The stability boundaries are the result of a competition between the damping, which radially shrinks the Wigner function, and the squeezing and dephasing, which are analyzed in depth here. Ignoring the constants in the Hamiltonians, the Liouvillians $\L_s$ with $s=\down,\up$ can be rewritten as
\begin{subequations}\label{eq:Lappx}
\begin{align}
    &\L_s = -\mri \Big[\hH^{\rm (rot)}_s + \hH^{\rm (sqz)}_s,\,\,\cdot\,\,\Big] + \kappa(1-\gamma)\D[\ha] + \kappa\gamma\D[\hn], \qq{where} \\
    &\hH^{\rm (rot)}_\down = \omega_0 (1-\frac{g^2}{2})\hn, \quad
    \hH^{\rm (sqz)}_\down  = -\frac{\omega_0 g^2}{4}(\ha^2+\ha^{\dagger2}),\\
    &\hH^{\rm (rot)}_\up   = \omega_0 (1+\frac{g^2}{2})\hn, \quad
    \hH^{\rm (sqz)}_\up    = +\frac{\omega_0 g^2}{4}(\ha^2+\ha^{\dagger2}).
\end{align}
\end{subequations}

The Hamiltonian term $\hH^{\rm (sqz)}_s$ represents squeezing, which stretches the Wigner function along one direction and shrinks it along the other orthogonal direction in the phase space. A convenient choice of coordinates to describe the effects of the squeezing term is 
\begin{align}
    \hat{x}'&= \frac{1}{\sqrt{2}}(\ha e^{\mri\pi/4} + \ha^\dagger e^{-\mri \pi/4}) = \hat{x}\cos\frac{\pi}{4} - \hat{p}\sin\frac{\pi}{4}, \\
    \hat{p}' &= \frac{1}{\sqrt{2}}(\ha e^{-\mri\pi/4} + \ha^\dagger e^{\mri \pi/4}) = \hat{x}\cos\frac{\pi}{4} + \hat{p}\sin\frac{\pi}{4},
\end{align}
such that
\begin{equation}
    -(\ha^2 + \ha^{\dagger 2}) = \hat{x}'\hat{p}' + \hat{p}'\hat{x}'.
\end{equation}
Then, the evolution of the Wigner function $W(x', p')$ with respect to the squeezing term $-(\ha^2 + \ha^{\dagger 2})$ is given by~\cite{Walls_Milburn_2008}
\begin{align}
    \partial_t W &= \lbrace\!\lbrace x'p'+p'x', \: W \rbrace\!\rbrace = \{ x'p'+p'x', \: W\} \label{eq:braket}\\
    &= p'\partial_{p'}W - x'\partial_{x'}W, \label{eq:squeezeWigner}
\end{align}
where $\lbrace\!\lbrace \cdot \rbrace\!\rbrace$ is the Moyal bracket and $\{\cdot\}$ is the Poisson bracket. The second equality in Eq.~\eqref{eq:braket} holds because the squeezing Hamiltonian is quadratic in $(x',p')$~\cite{hillery1984distribution}. The differential equation~\eqref{eq:squeezeWigner} clearly shows that, in the spin-$\down$ sector, the Wigner function is stretched in the $x'$ direction and shrunk in the $p'$ direction, where the rate of stretch (shrinkage) increases with the magnitude of $x'$ ($p'$). In the spin-$\up$ sector, the roles of $x'$ and $p'$ are interchanged. 

The Hamiltonian term $\hH^{\rm (rot)}_s$ represents rotation of the Wigner function. Indeed, $\hn = \frac{1}{2}(\hat{x}^2 + \hat{p}^2)-\frac{1}{2}$ such that the evolution of $W(x, p)$ with respect to $\hH^{\rm (rot)}_s$ is determined by
\begin{equation}
    \partial_t W = \frac{1}{2}\lbrace\!\lbrace x^2 + p^2, \: W \rbrace\!\rbrace 
    = \frac{1}{2}\{x^2 + p^2, \: W\}
    = 2(x \partial_{p}W - p\partial_{x}W)
    = \partial_\theta W, \label{eq:rotateWigner}
\end{equation}
where $(x, p) = (r\cos\theta, r\sin\theta)$. Note that according to Eq.~\eqref{eq:Lappx}, the rotation is slower in the spin-$\down$ sector for any $g\neq 0$; that is,
\begin{equation}\label{eq:omegaeff}
    |\omega^{\rm (eff)}_\down| < |\omega^{\rm (eff)}_\up|, \qq{where} 
    \omega^{\rm (eff)}_\down := \omega_0 (1-\frac{g^2}{2}), \quad 
    \omega^{\rm (eff)}_\up := \omega_0 (1+\frac{g^2}{2})
\end{equation}
and $\omega^{\rm (eff)}_\down = 0$ for $g=\sqrt{2}$. 
\begin{figure}[t]
\includegraphics[width=\textwidth]{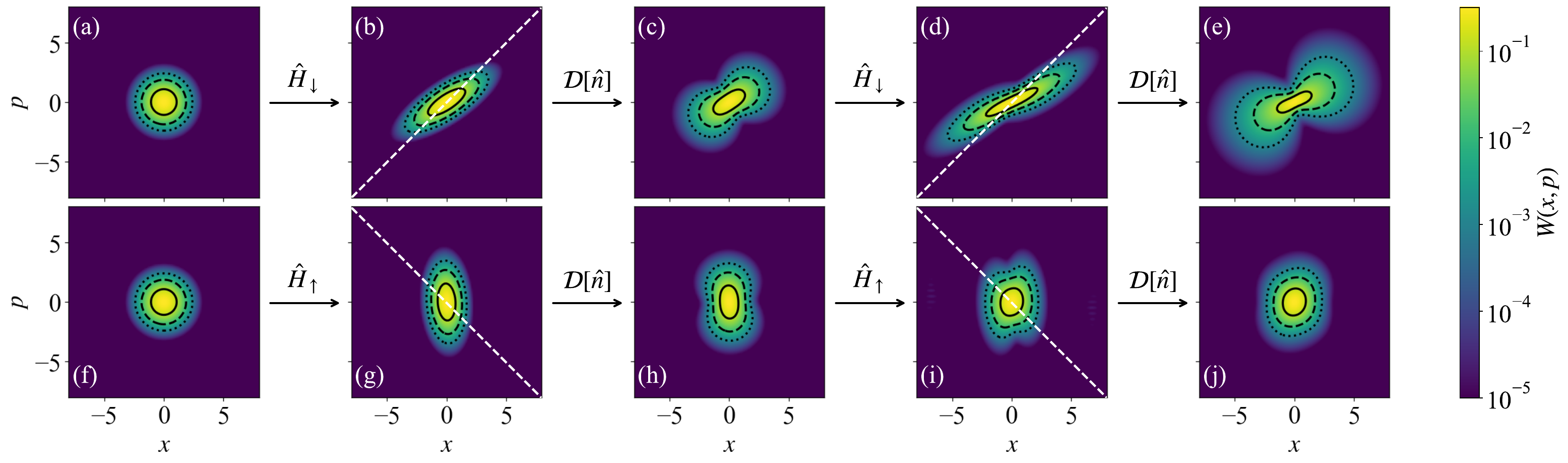}
	\caption{\label{fig:wigner} Trotterized evolution of the Wigner function with respect to $\L_\down$ in panels \textbf{(a)} to \textbf{(e)} and with respect to $\L_\up$ in panels \textbf{(f)} to \textbf{(j)}, starting from the vacuum state. Here we consider the case of pure dephasing ($\gamma=1$). The white dashed lines represent the direction along which the Wigner function is stretched by the squeezing Hamiltonian. The solid, dashed, and dotted contour lines represent the values $10^{-1}$, $10^{-2}$, and $10^{-3}$, respectively. Parameter values are set as $g=1.05$ and $\kappa/\omega_0 = 0.1$, and the evolution time between consecutive columns is $\Delta t =1/\omega_0$.}
\end{figure}

Using Bopp operators \cite{hillery1984distribution}, one finds that the evolution due to the damping term $\mathcal{D}[\ha]$ is
\begin{equation}
    \partial_t W = \frac{1}{2}\left[\partial_x(xW)+\partial_p(pW)\right]+\frac{1}{4}\left[\partial_x^2 W+\partial_p^2W\right].
\end{equation}
The first drift (friction) term can be written as $\nabla\cdot(\vec{r}W)$, i.e., causes radial contraction, and the second term $\nabla^2 W/4$ causes isotropic diffusion, both of which balance each other out in the ground state.

Finally, the dephasing term $\mathcal{D}[\hn]$ represents angular diffusion. The Wigner function evolves with respect to $\mathcal{D}[\hn]$ as
\begin{equation}
    \partial_t W = \frac{1}{4}\lbrace\!\lbrace x^2 + p^2, \: \lbrace\!\lbrace x^2 + p^2, \: W \rbrace\!\rbrace \rbrace\!\rbrace
    =\frac{1}{4} \{ x^2 + p^2, \: \{ x^2 + p^2, \: W \} \}
    \stackrel{\eqref{eq:rotateWigner}}{=} \partial_\theta^2 W.
\end{equation}
This is a diffusion equation for the angular direction. Thus, dephasing by itself does not change the radial moments $\langle (\hat{x}^2 + \hat{p}^{2})^k \rangle \sim \langle \hn^k \rangle$. However, we will see below that it can lead to a growth of radial moments in the interplay with squeezing and rotation. 

(i) The Wigner function dynamics can qualitatively explain why $g_{c,\down}^{(k)} \leq g_{c,\up}^{(k)}$ as rigorously shown in Appx.~\ref{appx:odd-k_up}. The top (bottom) row of Fig.~\ref{fig:wigner} visualizes the Trotterized evolution of the Wigner function with respect to $\L_\down$ ($\L_\up$), starting from the vacuum state. The white dashed lines in the top (bottom) row represent the $x'$ ($p'$) direction, along which the Wigner function is stretched by the squeezing Hamiltonian $\hH^{\rm (sqz)}_s$. From the first to second column, the Wigner function is both squeezed and rotated, where the rotation angle is larger in the spin-$\up$ sector (bottom row) due to the larger angular frequency [Eq.~\eqref{eq:omegaeff}]. This is followed by angular diffusion due to dephasing in the third column. Crucially, during the next Hamiltonian evolution between the third and fourth columns, the Wigner function in the spin-$\up$ sector (bottom row) is less stretched, as the density that had been stretched out during the previous Hamiltonian evolution was more effectively rotated away from the stretch axis by $\hH^{\rm (rot)}_\up$. As the stretch rate increases with the distance from the origin along the stretch axis [Eq.~\eqref{eq:squeezeWigner}], fast rotation suppresses stretching. The following dephasing in the last column clearly shows that, in the spin-$\up$ sector, the radial spread of the Wigner function occurs to a lesser extent than in the spin-$\down$ sector, explaining why observables diverge in the spin-$\up$ sector at larger $g$ values than in the spin-$\down$ sector. 

(ii) For a stable oscillator (any $g$ in the spin-$\up$ sector; $g<1$ for spin-$\down$), energy contours in phase space are elliptic. Hence, in addition to reducing quantum coherence, the angular diffusion due to dephasing moves weight into higher-energy regions, i.e., it has a heating effect. This heating, which has to be kept in check by damping, effectively moves weight into the tails of the Wigner function. This explains how dephasing can contribute to the divergence of (higher-order) observables and why we see instabilities for infinitesimal dephasing at $\gamma=1$ (no damping).
\begin{figure}[t]
\includegraphics[width=0.95\textwidth]{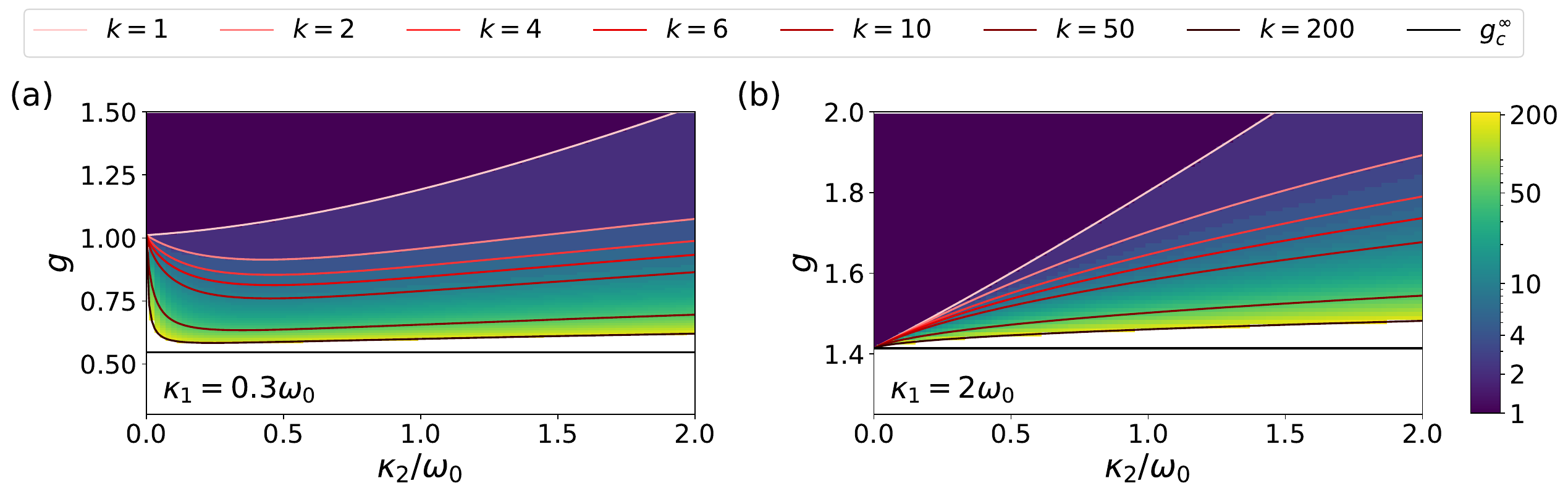}
	\caption{\label{fig:k2gboundaries} Stability boundaries of the $k$th-order bosonic operators for the spin-$\down$ sector in the thermodynamic limit $\eta\rightarrow\infty$, shown in the $\kappa_2$--$g$ plane, where $\kappa_2 := \kappa\gamma$ is the dephasing strength. The damping strength $\kappa_1:=\kappa(1-\gamma)$ is fixed to  \textbf{(a)} $\kappa_1=0.3\omega_0$ and \textbf{(b)} $\kappa_1=2\omega_0$. Panel (b) represents an exceptional case where $g_c=g_c^\infty=\sqrt{2}$. Note that, for fixed $\kappa_1$, $g_c^\infty = \sqrt{\kappa(1-\gamma)/\omega_0} = \sqrt{\kappa_1/\omega_0}$ is constant.}
\end{figure}

(iii) In line with its interpretation as a source of heating, (small) dephasing generally exacerbates the divergence of the bosonic operators' steady-state expectation values: 

If we just decreased damping linearly as $1-\gamma$ and \emph{not} 
simultaneously ramp up the dephasing, we would see the stability boundary $\sqrt{1+(y/2)^2}$ for all operator orders $k$ according to Eq.~\eqref{eq:MFT-gc}, where $y:=\kappa(1-\gamma)/\omega_0$. This is strictly larger than the $k\to\infty$ stability boundary $\sqrt{y}$ with dephasing [Eq.~\eqref{eq:gcinfty}] except for the exceptional point $y=2$.

Furthermore, starting from the Gaussian critical point $(\gamma, g) = (0, g_c)$, when small dephasing is added, the $k$th-order ($k\geq 2$) operators immediately diverge at a smaller $g<g_c$ with an intriguing exception for $\kappa/\omega_0=2$, where $g_c=g_c^\infty(\gamma=0)=\sqrt{2}$. To show this concretely, we define the damping strength $\kappa_1$ and the dephasing strength $\kappa_2$ as
\begin{equation}
    \kappa_1 := \kappa(1-\gamma) \qq{and} \kappa_2 := \kappa\gamma,
\end{equation}
and draw the stability boundaries in a $\kappa_2$--$g$ plane with fixed damping $\kappa_1$ in Fig.~\ref{fig:k2gboundaries}a. Starting from the Gaussian critical point $(\kappa_2, g) = (0, g_c)$, small dephasing $\kappa_2$ lowers the $k\geq 2$ stability boundaries towards $g_c^\infty$. 
Figure~\ref{fig:k2gboundaries}b shows the exceptional case $\kappa_1/\omega_0=2$ where $g_c = g_c^\infty$ and the stability boundaries are monotonically increasing with respect to $\kappa_2$. 

This can also be qualitatively understood from the Wigner function dynamics. Squeezing is the only operation that actually pushes weight into the tail of the Wigner function along the stretch axis and can hence grow the magnitude of observables such as radial moments. In the Trotterized evolution as in Fig.~\ref{fig:wigner}, after the Wigner function is stretched and rotated, the angular diffusion due to small dephasing brings back some of the quasi-probability to the stretch axis. This causes the Wigner function to stretch further in the following squeezing step, as the rate of growth is proportional to the magnitude of the coordinate along the stretch axis, resulting in faster growth of radial moments, which is consistent with the trend in Fig.~\ref{fig:k2gboundaries}a where finite-$k$ stability boundaries are lowered as $\kappa_2$ increases from zero.
An exception occurs when rotation is absent as, in this case, dephasing only diffuses the distribution away from the stretch axis and therefore suppresses the growth of radial moments. This is consistent with the opposite trend in Fig.~\ref{fig:k2gboundaries}b for $\kappa_1/\omega_0=2$ where, indeed, rotation is absent in the spin-$\down$ sector ($\omega^{\rm (eff)}_\down = 0$).

In Fig.~\ref{fig:k2gboundaries}a, for larger $\kappa_2$, the finite-$k$ stability boundaries pass a minimum and start rising with respect to $\kappa_2$. In this regime, diffusion and the associated angular redistribution of quasi-probability have become so strong that, instead of the compensating effect on rotation, they actually contract the projection of the quasi-probability on the stretch axis compared to smaller $\kappa_2$.

Note that the exceptional case in Fig.~\ref{fig:k2gboundaries}b occurs when the Gaussian critical point $g_c$ and the non-Gaussian critical point $g_c^\infty$ coincide at $\gamma=0$. This points to an interesting connection between the Wigner-function picture in Fig.~\ref{fig:wigner}, where rotation exists except when $g=\sqrt{2}$, and the normal-phase boundary shown in the inset of Fig.~\ref{fig:boundary}, which is discontinuous at $\gamma=0$ except when $g_c=\sqrt{2}$.

\section{Small \texorpdfstring{$\gamma$}{gamma} perturbation theory for \texorpdfstring{$\hH^{(k)}$}{Hk}}\label{appx:PT}
In this appendix, we conduct the first-order perturbation theory to analyze the spectrum of the non-Hermitian spin-$k/2$ Hamiltonians $\hH^{(k)}$ for small $\gamma$. The result is used in the main text to assess the cascade of instabilities at small $\gamma$.

We first decompose $\hH^{(k)}$ from Eq.~\eqref{eq:eff_Hk_appx} into unperturbed and perturbed parts as
\begin{align}
    \hH^{(k)}=\hH^{(k)}_0+\hH^{(k)}_1 \qq{with}
    \hH^{(k)}_0=-\frac{\kappa k}{2}(1-\gamma)-g^2\omega_0\hS_x+(g^2-2)\mri \omega_0\hS_y  \qq{and }
    \hH^{(k)}_1=-2\kappa\gamma\hS_y^2.
\end{align}
Using the identity $e^{-t\hS_z}\hS_x e^{t\hS_z}=\cosh t\, \hS_x-\mri\sinh t\,\hS_y$, the unperturbed part can be transformed into
\begin{align}
    e^{t\hS_z}\hH^{(k)}_0 e^{-t\hS_z}=-\frac{\kappa k}{2}(1-\gamma)-2\omega_0\sqrt{g^2-1}\hS_x \qq{with }
    t=\sinh^{-1}\left(\frac{g^2-2}{2\sqrt{g^2-1}}\right).
\end{align}
The ``ground state'' of $e^{t\hS_z}\hH^{(k)}_0 e^{-t\hS_z}$, whose eigenvalue has the largest real part, is fully polarized along the $-x$ direction, i.e., the state $\ket{\leftarrow}$ with $\hS_x\ket{\leftarrow}=-S\ket{\leftarrow}$ and $S=k/2$. It has the $\hH^{(k)}_0$ eigenvalue
\begin{equation}
    \ell^{(k)}_{\text{max},0}=-\frac{\kappa k}{2}(1-\gamma)+\omega_0 k\sqrt{g^2-1}.
\end{equation}
The first-order correction due the perturbation $\hH^{(k)}_1$ is
\begin{align}
    \delta \ell^{(k)}_{\text{max}}&=\bra{\leftarrow} e^{t\hS_z} \hH^{(k)}_1 e^{-t\hS_z}  \ket{\leftarrow}  \nonumber \\
    &=-2\kappa\gamma\bra{\leftarrow} e^{t\hS_z} \hS_y e^{-t\hS_z} e^{t\hS_z}\hS_y e^{-t\hS_z} \ket{\leftarrow}\nonumber \\
    &=-2\kappa\gamma\bra{\leftarrow} (\cosh t\,\hS_y+\mri  \sinh t\,\hS_x)^2 \ket{\leftarrow} \nonumber \\
    &=-2\kappa\gamma\bra{\leftarrow} (\mri\cosh t\,\hS_x^+-\mri\cosh t\,\hS_x^-+\mri \sinh t\,\hS_x )^2 \ket{\leftarrow} \nonumber \\
    &=-2\kappa\gamma\bra{\leftarrow} (\cosh^2t\, \hS_x^-\hS_x^+ -\sinh^2t\,\hS_x^2) \ket{\leftarrow} \nonumber \\
    &=-2\kappa\gamma\frac{2 Sg^4- S^2(g^2-2)^2}{4(g^2-1)}.
\end{align}
Finally, inserting $S=k/2$, we obtain the largest real-part eigenvalue of $\hH^{(k)}$ to leading order in $\gamma$,
\begin{equation}
    \ell^{(k)}_{\text{max}} = \ell^{(k)}_{\text{max},0}+\delta \ell^{(k)}_{\text{max}}
    = k\left( -\frac{\kappa}{2}+\omega_0\sqrt{g^2-1}+\frac{\kappa\gamma(k-1)(g^2-2)^2}{8(g^2-1)} \right) + \order{\gamma^2}.
\end{equation}

\section{Zero-\texorpdfstring{$\eta$}{eta} analysis}\label{appx:zeroeta}
In this work, we mainly consider the soft-mode limit $\eta\to\infty$, which coincides with the thermodynamic limit and features a cascade of instabilities. In this appendix, we shall consider the opposing regime of vanishing $\eta=\Omega/\omega_0$, and we will find that the non-Gaussian open Rabi model is then stable for all $\gamma < 1$ regardless of the spin-boson coupling strength $\lambda$. This supports the hypothesis that the model is stable for any finite $\eta$, which is further substantiated by the mean-field theory, numerical data, and scaling analysis of the Keldysh action in Appx.~\ref{appx:Keldysh}. 

For $\eta=\Omega/\omega_0$ ($\Omega\to 0$), states with the qubit in $\hsigma^x$ eigenstates $\ket{\Right}$ and $\ket{\leftarrow}$ evolve independently. Taking for $\Omega=0$ the projection of $\hH_\text{Rabi}$ onto the $+1$ eigenstate $\ket{\Right}$ yields the reduced Hamiltonian
\begin{equation}
	\hH_\Right=\lambda(\ha+\ha^\dag)+\omega_0\hn
\end{equation}
for the oscillator. In particular, the term $-\frac{\omega_0 g^2}{4}(\ha+\ha^\dag)^2$ of the $\eta\to\infty$ spin-$\down$ Hamiltonian \eqref{eq:Hdown}, which includes squeezing $\sim(\ha^2+\ha^{\dag 2})$, is replaced by the coherent driving term $\lambda(\ha+\ha^{\dag})$ in $\hH_\Right$. The EOM for the $k$th-order bosonic Green's functions still form a closed hierarchy. For example, up to order $k=2$, the evolution of bosonic operators is determined by
\begin{subequations}
\begin{align}
	\L_\Right^\dag(\ha) &= -\left(\mri\omega_0+\frac{\kappa}{2}\right)\ha-\mri\lambda,\\
	\L_\Right^\dag(\ha^\dag\ha) &= \mri\lambda(\ha-\ha^\dag)-(1-\gamma)\kappa\ha^\dag\ha,\qq{and} \\
	\L_\Right^\dag(\ha^2) &=-2\mri\lambda\ha -\big(2\mri\omega_0 + (1+\gamma)\kappa\big)\ha^2,
\end{align}
\end{subequations}
where the Liouvillian $\L_\Right$ is given by Eq.~\eqref{eq:mastereqn2} with $\hH_\down$ replaced by $\hH_\Right$. The coherent driving only couples the $k$th-order Green's functions to the order-$(k-1)$ Green's function. The non-Hermitian spin-$k/2$ Hamiltonian $\hH^{(k)}_\Right$ for the $k$-th diagonal block in the generator of the Green's function EOM can be obtained by simply setting $g=0$ in Eq.~\eqref{eq:eff_Hk_appx}. The resulting
\begin{equation}
	\hH^{(k)}_\Right=-\frac{\kappa k}{2}(1-\gamma)-2\mri \omega_0\hS_y-2\kappa\gamma\hS_y^2
\end{equation}
is diagonal in the $\hS_y$ eigenbasis with the $k+1$ eigenvalues
\begin{equation} \label{eq:zeroeta}
	\ell^{(k)}_{M_y}= -\frac{\kappa k}{2}(1-\gamma)-2\mri \omega_0 M_y-2\kappa\gamma M_y^2,\qq{where}
	M_y=-\frac{k}{2},-\frac{k}{2}+1,\dotsc,\frac{k}{2}.
\end{equation}

Also, $\hH^{(k)}_\Left$ is obtained by replacing $\lambda\to -\lambda$. As $\hH^{(k)}_\Right$ is already $\lambda$ independent, both effective Hamiltonians agree and, in the $\eta\to 0$ limit, the full system is stable as long as $\gamma<1$ because
\begin{equation}
	\text{Re}(\ell^{(k)}_{M_y})\leq-\frac{\kappa k}{2}(1-\gamma)\leq 0\quad\text{for all}\ \gamma\in[0,1].
\end{equation}

\section{Degenerate perturbation theory for \texorpdfstring{$\expect{\hsigma^z}_s$}{the z magnetization}}\label{appx:sigmaz}
Let us now determine the steady-state expectation value of the qubit observable $\hsigma^z$ to leading order in $1/\eta$ as stated in Eq.~\eqref{eq:sigmazs}. While, for any finite $\eta$, we have a unique steady state $\dmf_s(\eta)$, the qubit $\down$ and $\up$ states decouple in the soft-mode limit $\eta\to\infty$, giving us a two-dimensional steady-state manifold. In the spirit of zeroth-order degenerate perturbation theory, we want to determine
\begin{equation}\label{eq:ssansatz}
    \dmf_s:= \lim_{\eta\to \infty} \dmf_s(\eta) = p_\down \ket{\down}\bra{\down} \otimes \dm_\down + p_\up \ket{\up}\bra{\up} \otimes \dm_\up,
\end{equation}
which is a convex combination of the $\eta\to\infty$ steady states for spin $\down$ and $\up$, i.e.,
$p_\down, p_\up \geq 0$ and $p_\down + p_\up = 1$ with $\dm_\down$ and $\dm_\up$ being the steady states of $\L_\down$ [Eq.~\eqref{eq:mastereqn2}] and $\L_\up$ [Eq.~\eqref{eq:mastereqn2-up}], respectively.

The leading $1/\eta$ correction to the Schrieffer-Wolff transformed Rabi Hamiltonian $\hU^\dag \hH_\text{Rabi} \hU$ from Eq.~\eqref{eq:HafterSW} is $-\frac{\omega_0 g^4}{16\eta}\hsigma^z (\ha+\ha^\dag)^4$, where $\hU = \exp[-(g/2\sqrt{\eta})(\ha+\ha^\dag)(\hsigma^+ - \hsigma^-)]$ \cite{hwang2015quantum}. Up to order $1/\eta$, the transformed dissipators are
\begin{alignat}{4}\label{eq:Dp-a}
    \D'[\ha] &:= \D[\hU^\dag\ha\hU] 
    &&= \D\left[\ha - \frac{\mri g}{2\sqrt{\eta}}\hsigma^y\right]\qq{and}\\
    \label{eq:Dp-n}
    \D'[\ha^\dag\ha] &:= \D[\hU^\dag\ha^\dag\ha\hU]
    &&= \D\left[\ha^\dag\ha + \frac{\mri g}{2\sqrt{\eta}}\hsigma^y(\ha - \ha^\dag)\right].
\end{alignat}
Thus, up to order $1/\eta$, the transformed Liouvillian of the open quantum Rabi model is
\begin{equation}\label{eq:L-etaExpand}
    \L' := -\mri \left[\frac{\Omega}{2}\hsigma^z + \frac{\omega_0g^2}{4}\hsigma^z(\ha+\ha^\dag)^2+\omega_0\ha^\dag\ha-\frac{\omega_0 g^4}{16\eta}\hsigma^z (\ha+\ha^\dag)^4,\:\cdot\:\right]+\kappa(1-\gamma)\D'[\ha]+\kappa\gamma\D'[\ha^\dag\ha].
\end{equation}

The steady-state expectation value $\expect{\hsigma^z}_s := \Tr(\hsigma^z \dmf_s)$ obeys the equation
\begin{align}\nonumber
	0 &= \partial_t\Tr(\hsigma^z \dmf_s) = \Tr(\hsigma^z \partial_t \dmf_s) = \Tr(\hsigma^z \L'(\dmf_s)) \\ 
	&= (1-\expect{\hsigma^z}_s) \Tr\left[\hsigma^z \L'(\ket{\down}\bra{\down} \otimes \rho_\down)\right] + (1+\expect{\hsigma^z}_s) \Tr\left[ \hsigma^z \L'(\ket{\up}\bra{\up} \otimes \rho_\up)\right],
	\label{eq:selfconsistency}
\end{align}
where we have used that $p_\down = (1-\expect{\hsigma^z}_s)/2$ and $p_\up = (1+\expect{\hsigma^z}_s)/2$.

With straightforward calculations, the traces in Eq.~\eqref{eq:selfconsistency} evaluate to 
\begin{subequations}\label{eq:downterm}
\begin{align}
    \Tr\left[\hsigma^z \L'(\ket{\down}\bra{\down} \otimes \rho_\down)\right]
    &= \kappa(1-\gamma)\frac{g^2}{2\eta} - \kappa\gamma \frac{g^2}{2\eta} \Tr\left[ (\ha-\ha^\dag)^2 \rho_\down\right]\qq{and} \\
    \Tr\left[\hsigma^z \L'(\ket{\up}\bra{\up} \otimes \rho_\up)\right]
    &= -\kappa(1-\gamma)\frac{g^2}{2\eta} + \kappa\gamma \frac{g^2}{2\eta} \Tr\left[ (\ha-\ha^\dag)^2 \rho_\up\right],
\end{align}
\end{subequations}
where we use $\L'\rvert_{\eta=\infty}(\ket{\down}\bra{\down} \otimes \rho_\down)=\L'\rvert_{\eta=\infty}(\ket{\up}\bra{\up} \otimes \rho_\up)=0$.
Plugging Eqs.~\eqref{eq:downterm} into Eq.~\eqref{eq:selfconsistency}, and defining
\begin{equation}\label{eq:ydownyup}
     y_\down := \Tr[(\ha-\ha^\dag)^2 \dm_\down],\quad
     y_\up := \Tr[(\ha-\ha^\dag)^2 \dm_\up],
\end{equation}
we obtain the equation
\begin{equation}
    \kappa(1-\gamma)\frac{g^2}{\eta}\expect{\hsigma^z}_s + \kappa\gamma \frac{g^2}{2\eta} \Big( (1-\expect{\hsigma^z}_s) y_\down - (1+\expect{\hsigma^z}_s) y_\up \Big) = 0.
\end{equation}
Solving for $\expect{\hsigma^z}_s$, we arrive at the result
\begin{equation}\label{eq:sigmazs-appx}
     \expect{\hsigma^z}_s = \frac{\gamma(y_\down - y_\up)}{-2(1-\gamma) + \gamma (y_\down + y_\up)}.
\end{equation}

We conclude this appendix by determining the parameters $y_\down$ and $y_\up$. For the $\down$-state projection $\dm_\down$, the linear steady-state equations $\partial_t\vec{v}_2 = H_2\vec{v}_2+Y_2=0$ for the vector $\vec{v}_2 := (\expect{\ha\ha}, \expect{\ha^\dag\ha}, \expect{\ha^\dag\ha^\dag})^T$ of second-order bosonic operators
from Eq.~\eqref{eq:k2EOM} have the solution
\begin{equation}
\begin{pmatrix}
    \expect{\ha\ha}_s \\
    \expect{\ha^\dag\ha}_s \\
    \expect{\ha^\dag\ha^\dag}_s
\end{pmatrix}
= \frac{1}{\mc{N}_\down} 
\begin{pmatrix}\label{eq:k2EOMsoln}
    (1-\gamma)g^2 \big( 2-g^2 + \mri \frac{\kappa}{\omega_0}(1+\gamma) \big) \\
    (1+\gamma)g^4 \\
    (1-\gamma)g^2 \big( 2-g^2 - \mri \frac{\kappa}{\omega_0}(1+\gamma) \big)
\end{pmatrix},
\end{equation}
where the normalization constant is given by
\begin{equation}\label{eq:Ndown}
    \mc{N}_\down = 8(1-\gamma)\left( 1 + (1+\gamma)^2 \frac{\kappa^2}{4\omega_0^2}\right) - 8(1-\gamma)g^2 - 4\gamma g^4.
\end{equation}
One can check that $\mc{N}_\down = 0$ at $g=g^{(2)}_c$, which is consistent with the fact that $g=g^{(2)}_c$ is the stability boundary of the second-order bosonic operators. 
With the solution \eqref{eq:k2EOMsoln}, the parameter $y_\down$ from Eq.~\eqref{eq:ydownyup} evaluates to
\begin{equation}\label{eq:ydown}
    y_\down = \expect{\ha\ha}_s + \expect{\ha^\dag\ha^\dag}_s - 2\expect{\ha^\dag \ha}_s - 1 = \frac{4g^2}{\mc{N}_\down}(1-\gamma-g^2) - 1. 
\end{equation}

For the $\up$-state projection $\dm_\up$, we recall that $\hH_\up$ and $\L_\up$ are obtained from $\hH_\down$ and $\L_\down$ by the substitution $g^2 \to -g^2$. With the same substitution in Eqs.~\eqref{eq:ydown} and \eqref{eq:Ndown}, we obtain
\begin{align}\label{eq:yup}
    y_\up &= -\frac{4g^2}{\mc{N}_\up}(1-\gamma+g^2) - 1\qq{with}\\
    \mc{N}_\up &= 8(1-\gamma)\left( 1 + (1+\gamma)^2 \frac{\kappa^2}{4\omega_0^2}\right) + 8(1-\gamma)g^2 - 4\gamma g^4.
\end{align}
Note that $|y_{\up}|$ is finite at $g=g^{(2)}_c$ as $\mc{N}_\up>0$. This is due to the ordering $g^{(2)}_c\equiv g^{(2)}_{c,\down}\leq g^{(2)}_{c,\up}$ of the two corresponding stability boundaries as discussed in Appx.~\ref{appx:odd-k_up}. 

With Eqs.~\eqref{eq:sigmazs-appx}, \eqref{eq:ydown}, and \eqref{eq:yup}, we have determined the steady-state expectation value $\lim_{\eta_\to \infty}\expect{\hsigma^z}_s$ for the soft-mode limit in the $g < g^{(2)}_c$ regime as stated in Eq.~\eqref{eq:sigmazs} of the main text. This analytical result is shown by solid lines in Fig.~\ref{fig:etascaling}a.

\section{Convergence with respect to \texorpdfstring{$N_{\rm max}$}{Nmax} and Fock distribution}\label{appx:Nmax}
\begin{figure}[t]
	\includegraphics[width=0.6\linewidth]{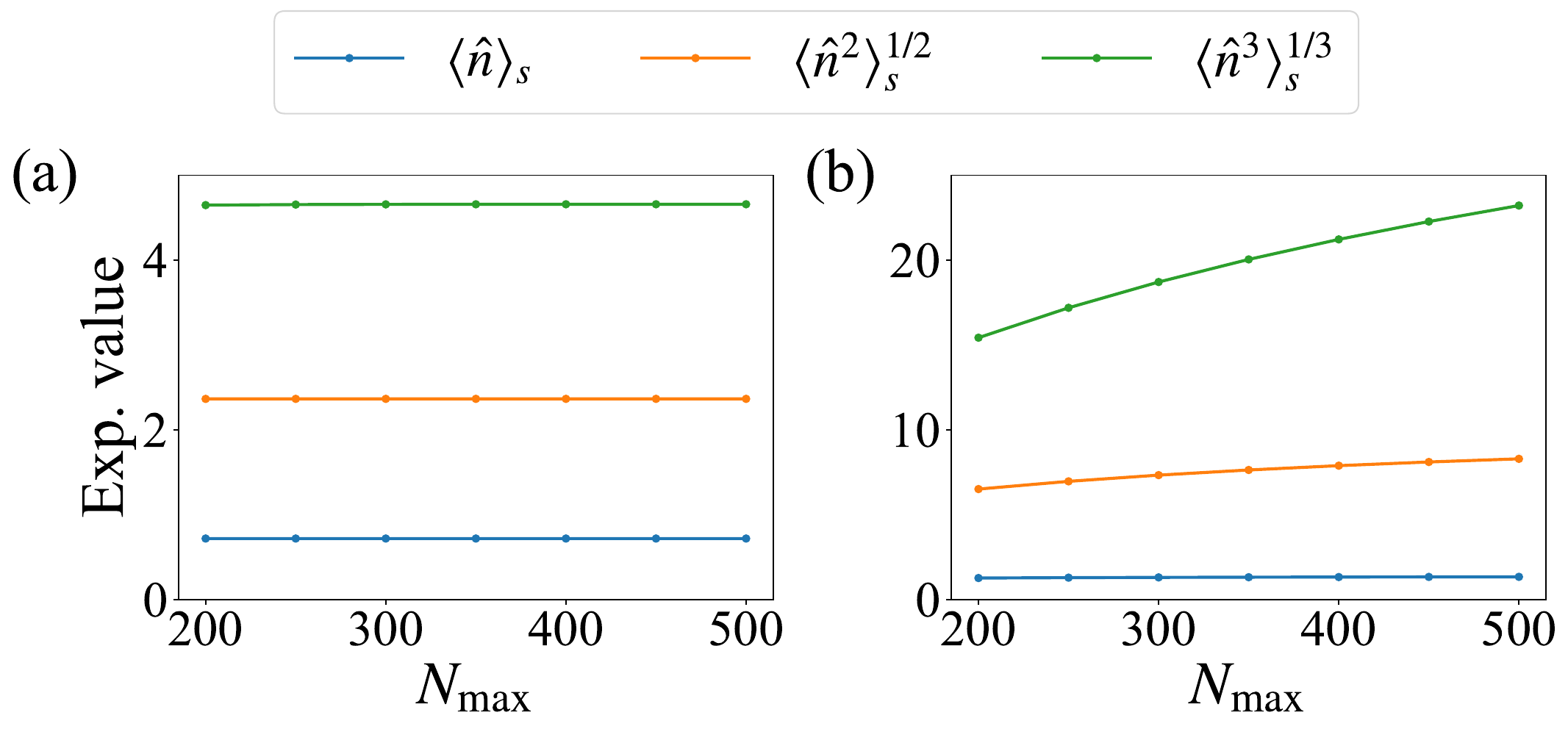}
	\caption{\label{fig:Nmax} Steady-state expectation values for the 2nd, 4th, and 6th-order operators $\hn$, $\hn^2$, and $\hn^3$ at $\kappa/\omega_0=1$, $\eta=10^4$, $g=1.05$ for \textbf{(a)} $\gamma=0.1$ such that $g<g_c^{(6)}$ and \textbf{(b)} $\gamma=0.3$ such that $g_c^{(4)}<g<g_c^{(2)}$. }
\end{figure}

Supplementing Fig.~\ref{fig:hierarchy}, Fig.~\ref{fig:Nmax} provides data for the dependence of the bosonic observables $\expect{\hn}_s$, $\expect{\hn^2}_s$, and $\expect{\hn^3}_s$ on the number of retained oscillator energy levels $N_{\rm max}$. Here we choose $\eta=10^4$, which is the largest value considered in Fig.~\ref{fig:hierarchy}. For $\gamma=0.1$, all 3 expectation values converge, which is consistent with $g < g_c^{(6)}$. For $\gamma=0.3$, $\expect{\hn}_s$ converges, but $\expect{\hn^2}_s$ and $\expect{\hn^3}_s$ continue to grow with $N_{\rm max}$, which is consistent with $g_c^{(4)}<g<g_c^{(2)}$. Together with Fig.~\ref{fig:hierarchy}, these data support the cascade of instabilities predicted for the non-Gaussian DPT. 

\begin{figure}[t]
	\includegraphics[width=0.95\linewidth]{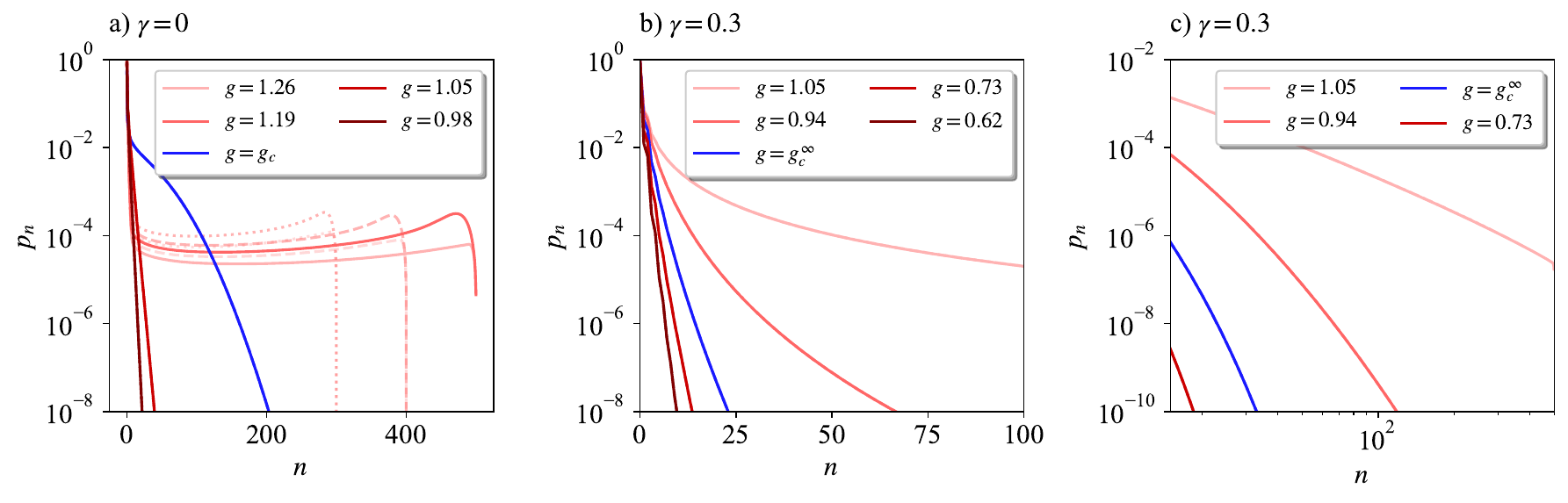}
	\caption{\label{fig:pn_dis} The Fock distribution $p_n$ for different system parameters and $N_{\rm max}$. Solid, dashed, and dotted lines represent $N_{\rm max}=500$, $400$, and $300$, respectively. In b) and c), lines with different $N_{\rm max}$ collapse almost perfectly. \textbf{(a)} $\gamma=0$, the Gaussian case. Here, $g_c \approx 1.118$. \textbf{(b, c)} $\gamma=0.3$, with linear $x-$axis and logarithmic $x-$axis, respectively. Here, $g_c^\infty \approx 0.837$.}
\end{figure}

The divergence of physical observables also manifests itself in the Fock distribution $p_n:=\Tr\left(\dmf\ket{n}\bra{n}\right)$. In Fig.~\ref{fig:pn_dis}, we plot the Fock distribution for both the Gaussian case ($\gamma=0$) and the non-Gaussian case ($\gamma=0.3$). For the Gaussian case, there is a universal critical point $g_c$ for observables of all orders. It can be clearly seen that for $g>g_c$, the Fock distributions do not decay in $n$ and do not converge with respect to $N_{\rm{max}}$ (the decay near $N_{\rm{max}}$ is an artifact of the numerical truncation), which signifies the dynamical instability. In contrast, for the non-Gaussian case shown in Figs.~\ref{fig:pn_dis}b and c, the Fock distributions are well-converged with respect to $N_{\rm max}$ and decay with increasing $n$. In this case, the divergence of observables like $\expect{\hn^3}_s$ for $g>g_c^{(3)}$ is due to the slow-decaying tail of the Fock distribution. The log-log plot in Fig.~\ref{fig:pn_dis}c shows that below $g_c^\infty$, the Fock distribution decays exponentially, guaranteeing the finiteness of expectation values. In contrast, the tail decays algebraically for $g>g_c^\infty$. Different system parameters give rise to different powers of this algebraic decay, resulting in the stability diagram in Fig.~\ref{fig:boundary2}. The algebraic decay of the Fock distribution is a clear signature for the non-Gaussian nature of the system with dephasing.

\section{Keldysh action and scaling dimensions}\label{appx:Keldysh}
Consider the reduced bosonic model~\eqref{eq:mastereqn2} for the soft-mode limit $\eta\to\infty$ and the qubit-$\down$ state. The corresponding partition function of the non-equilibrium steady state reads
\begin{equation}
    Z:=\lim_{t\to\infty} \Tr \dm(t) = \lim_{t\to\infty} \Tr\left[ (e^{\L t/N})^N\dm(0)\right]
    \stackrel{N\to\infty}{=}\int\mathscr{D}[\psi_{\pm},\psi_{\pm}^*]e^{-S}.
\end{equation}
In the first step, we have decomposed the evolution into $N$ Trotter steps. Inserting resolutions of the identity in terms of coherent states between the Trotter factors gives the partition function as a product of coherent-state matrix elements $\bra{\psi_+'}e^{\L\Delta t}\big[\ket{\psi_+}\bra{\psi_-}\big]\ket{\psi_-'}$ with $\ha\ket{\psi_\pm}=\psi_{\pm}\ket{\psi_\pm}$.
In the second step, taking the continuous-time limit $N\to\infty$, one arrives at the coherent-state path integral with the Keldysh action \cite{Kamenev2023,Sieberer2016-79}
\begin{align}
    S&=\int \ud t\,\Big[ \psi_+^*\partial_t\psi_+-\psi_-^*\partial_t\psi_- -\mathscr{L}(\psi_\pm,\psi_\pm^*) \Big] \nonumber\\
     &=\int \ud t\,\Big[ \psi_+^*\partial_t\psi_+ - \psi_-^*\partial_t\psi_-+\mri\omega_0\Big(|\psi_+|^2-|\psi_-|^2+\frac{g^2}{4}\left(\psi_-+\psi_-^*\right)^2-\frac{g^2}{4}\left(\psi_++\psi_+^*\right)^2\Big) \nonumber \\
    &\qquad\qquad -\kappa(1-\gamma)\Big(\psi_+\psi_-^*-\frac{1}{2}|\psi_+|^2-\frac{1}{2}|\psi_-|^2\Big)-\kappa\gamma \Big( |\psi_+|^2|\psi_-|^2-\frac{1}{2}|\psi_+|^4-\frac{1}{2}|\psi_-|^4 \Big) \Big],
\end{align}
where $\mathscr{L}(\psi_\pm,\psi_\pm^*)$ are the matrix elements of the Liouvillian.

After the Keldysh rotation with
\begin{align*}
    \psi_c:=\frac{1}{\sqrt{2}}(\psi_++\psi_-)\qq{and}
    \psi_q:=\frac{1}{\sqrt{2}}(\psi_+-\psi_-),
\end{align*}
the action attains the form
\begin{align}
    S&=\int \ud t\,\Big[\psi_c^*\partial_t\psi_q+\psi_q^*\partial_t\psi_c+\mri\omega_0\left(\psi_c\psi_q^*+\psi_q\psi_c^*\right)-\frac{\mri\omega_0g^2}{2}\left(\psi_c\psi_q+\psi_c^*\psi_q^*+\psi_c^*\psi_q+\psi_c\psi^*_q\right) \nonumber \\
    &\qquad\qquad- \frac{\kappa(1-\gamma)}{2}\left( \psi_q\psi_c^*-\psi_c\psi_q^*-2\psi_q\psi_q^*\right) +\frac{\kappa\gamma}{2}\left(\psi_c\psi_q^*+\psi_c^*\psi_q\right)^2  \Big],
    \label{eq:action}
\end{align}
where the quartic term $(\psi_c\psi_q^*+\psi_c^*\psi_q)^2$ is due to the dephasing. The quadratic part characterizes a standard squeezed oscillator with linear damping and no dephasing as studied in Ref.~\cite{hwang2018dissipative}.

We have found considerable discrepancies between mean-field theory and the actual physics of the non-Gaussian model. On the level of the Keldysh field theory \cite{Kamenev2023,Sieberer2016-79,Thompson2023-455}, this observation can be explained by determining engineering dimensions (a.k.a.\ canonical scaling dimensions) of the fields and coupling parameters for the vicinity of the Gaussian DPT at $(\gamma,g)=(0,g_c)$, elucidating the relevance of the quartic terms. 

Defining the scaling dimension of time as $[t]=-1$, using the invariance of the partition function under scaling transformations such that $[S]=0$, and noting that $\kappa$ should not scale in order to have the constant noise vertex $\sim\psi_q\psi_q^*$ in the action \cite{Sieberer2016-79}, the Gaussian parts yield the engineering dimensions
\begin{equation}
    [t]=-1,\quad[\kappa]=0,\quad [\psi_c]=-\frac{1}{2},\qq{and} [\psi_q]=\frac{1}{2}.
\end{equation}
which are consistent with corresponding analyses in Refs.~\cite{torre2013keldysh,hwang2018dissipative}.

For the quartic dephasing term in the action \eqref{eq:action} this implies
\begin{equation}
    [\gamma]=-[t]-[\kappa]-2([\psi_c]+[\psi_q])=1,
\end{equation}
indicating that dephasing is indeed a relevant perturbation for the model, in the sense that dephasing gets amplified through renormalization group transformations, driving us out of the vicinity of the Gaussian critical point. 

The relevance of finite-$\eta$ corrections can be assessed similarly. The corrections that contribute to the most relevant quartic terms like
\begin{equation}
    \frac{\alpha}{\eta}\int\ud t\, \psi_c\psi_c\psi_c\psi_q
\end{equation}
in the action, originate from the $1/\eta$ correction $-\frac{\omega_0 g^4}{16\eta}\hsigma^z (\ha+\ha^\dag)^4$ to the transformed Hamiltonian in Eq.~\eqref{eq:L-etaExpand} and $1/\eta$ terms in $\D'[\ha]\circ\D'[\ha^\dag\ha](\dm)$ from the transformed dissipators given in Eqs.~\eqref{eq:Dp-a} and \eqref{eq:Dp-n}. The corresponding engineering dimension is
\begin{equation}
    \left[\frac{\alpha}{\eta}\right]=-[t]-3[\psi_c]-[\psi_q]=2,
\end{equation}
which is also relevant, i.e., finite $\eta$ drives us away from the Gaussian DPT. Furthermore, this substantiates the hypothesis that the model is stable for any finite $\eta$, which is also supported by the mean-field theory, numerical data, and zero-$\eta$ analysis in Appx.~\ref{appx:zeroeta}.
With $[\eta]=-2$ and $[\expect{\ha^\dag\ha}]=[\psi_c^2]=-1$ at $(\gamma,g)=(0,g_c)$, we have the scaling
\begin{equation}
	\expect{\ha^\dag\ha}\propto \eta^{1/2}
\end{equation}
for the critical Gaussian model, which is consistent with the mean-field theory (Appx.~\ref{appx:MFT}) and the analysis in Ref.~\cite{hwang2018dissipative}.

With dephasing and finite-$\eta$ corrections, in the vicinity of the Gaussian critical point, the model features two relevant perturbations. In future work, the non-trivial renormalization group analysis could reveal the universal properties of the resulting non-Gaussian DPT.

\newpage
%

\end{document}